\documentclass[12pt]{article}
\usepackage{epsfig}
\usepackage{graphicx}

\usepackage{amssymb,amsmath}

\def\la{\langle }
\def\ra{ \rangle }
\def\la{\langle }

\newcommand{\beq}{\begin{equation}}
\newcommand{\eeq}{\end{equation}}
\newcommand{\bea}{\begin{eqnarray}}
\newcommand{\eea}{\end{eqnarray}}

\newcommand{\DL}{\mathcal{D}}
\newcommand{\PL}{\mathcal{P}}
\newcommand{\MTM}{\mathcal{M}}

\newcommand{\jpmdisclaimer}{\vskip3cm\par\noindent\textbf{\footnotesize
Information has been obtained from sources believed to be reliable
but JPMorgan Chase \& Co. or its affiliates and/or subsidiaries
(collectively JPMorgan) does not warrant its completeness or
accuracy.  Opinions and estimates constitute our judgement as of the
date of this material and are subject to change without notice.
Past performance is not indicative of future results.  This material
is not intended as an offer or solicitation for the purchase or sale
of any financial instrument.  Securities, financial instruments or
strategies mentioned herein may not be suitable for all investors.
The opinions and recommendations herein do not take into account
individual client circumstances, objectives, or needs and are not
intended as recommendations of particular securities, financial
instruments or strategies to particular clients. The recipient of
this presentation must make its own independent decisions regarding
any securities or financial instruments mentioned herein.  JPMorgan
may act as market maker or trade on a principal basis, or have
undertaken or may undertake account transactions in the financial
instruments or related instruments of any issuer discussed herein
and may act as underwriter, placement agent, advisor or lender to
such issuer. JPMorgan and/or its employees may hold a position in
any securities or financial instruments mentioned herein.  Important
disclosures are available on a company specific basis at
http://mm.jpmorgan.com/disclosures/company.  Additional information
available upon request.}}

\begin{document}
\begin{titlepage}
\begin{flushright}
\end{flushright}
\vskip3cm
\begin{center}
{\LARGE
BSLP: Markovian Bivariate Spread-Loss Model
\vskip0.5cm
for Portfolio Credit Derivatives}
\vskip1.0cm
{\Large Matthias Arnsdorf and Igor Halperin}
\vskip0.5cm
Quantitative Research, JP Morgan
\vskip0.5cm
Email: matthias.x.arnsdorf@jpmorgan.com, igor.halperin@jpmorgan.com
\vskip0.5cm
March 2007 \\
\vskip1.0cm
{\Large Abstract:\\}
\end{center}
\parbox[t]{\textwidth}{


BSLP is a two-dimensional dynamic model of interacting
portfolio-level loss and loss intensity processes. It is constructed
as a Markovian, short-rate intensity model, which facilitates fast
lattice methods for pricing various portfolio credit derivatives
such as tranche options, forward-starting tranches, leveraged
super-senior tranches etc. A semi-parametric model specification is
used to achieve near perfect calibration to any set of consistent
portfolio tranche quotes. The one-dimensional local intensity model
obtained in the zero volatility limit of the stochastic intensity is
useful in its own right for pricing non-standard index tranches  by
arbitrage-free interpolation. }

\vspace{2.0cm}
\newcounter{helpfootnote}
\setcounter{helpfootnote}{\thefootnote}
\renewcommand{\thefootnote}{\fnsymbol{footnote}}
\setcounter{footnote}{0} \footnotetext{ Opinions expressed in this
paper are those of the authors, and do not  necessarily reflect the
view of JP Morgan. We would like to thank Andrew Abrahams, Morten
Andersen, Anil Bangia, Rama Cont, Ian Dowker, Kay Giesecke, Dapeng
Guan, Regis Guichard, David Liu, Antonio Paras, Philipp
Sch{\"o}nbucher, Jakob Sidenius, Nicolas Victoir and Yulia
Voevodskaya for valuable discussions.
All remaining errors are our own.
}
\renewcommand{\thefootnote}{\arabic{footnote}}
\setcounter{footnote}{\thehelpfootnote}

 \end{titlepage}

\section{Introduction}

A large class of portfolio credit derivatives can be viewed as
derivatives referencing the cumulative portfolio losses.
Furthermore, we can distinguish between two classes of such
derivatives. For single-period (vanilla) instruments such as
synthetic CDO tranches all that is needed for pricing is the set of
marginal portfolio loss distributions at different time horizons.
More exotic, multi-period instruments such as  tranche options
require knowledge of the future distributions of the mark-to-market
(MTM) of the tranche as well as the portfolio losses. This is
equivalent to having a model for the evolution of the portfolio
losses, from which the MTM of the tranche can then be derived.

If we hedge such  multi-period products only using tranches on the
underlying portfolio, then the challenge of modeling single-name
dynamics does not arise at all. Therefore, under these conditions
both the pricing and risk management of portfolio credit derivatives
can be done within a model that specifies the portfolio-level
dynamics of cumulative losses but leaves the single name dynamics
unspecified (or specified at a later stage).

The fact that powerful and flexible pricing models can be
constructed within such a ``top-down'' framework\footnote{Here the
``top'' refers  to the portfolio-level dynamics, and the ``down''
refers to the single name dynamics which, if needed, might in
principle be constructed consistently with the top-level portfolio
dynamics. In this paper, we will only be concerned with the
portfolio-level dynamics.} was first recognized in work by Giesecke
and Goldberg~\cite{GG}, Sch{\"o}nbucher~\cite{Schon2005}, Sidenius,
Piterbarg and Andersen (SPA) \cite{SPA}, and Bennani \cite{Bennani}.
More recent publications along this strand include Brigo et
al.~\cite{Brigo}, Chapovsky et al.~\cite{Chapovsky}, Errais et
al.~\cite{EGG}, Cont et al.~\cite{Cont}, Ding et al.~\cite{Ding},
and de Kock et al.~\cite{Chains}. Furthermore, while this paper was
under preparation we learned of a recent paper by Lopatin and
Misirpashaev \cite{LM} who have independently suggested a model
similar to the one developed in this paper. We will comment on
similarities and differences between their approach and ours in due
course.

In this paper, we present the Bivariate Spread-Loss Portfolio (BSLP)
model - a dynamic Markovian model for correlated portfolio loss and
loss intensity processes. In our model the portfolio loss process
follows a Markov chain whose generator is driven by a stochastic
intensity (so that the generator itself becomes stochastic). The
intensity is given by a diffusion process which can incorporate
default-induced jumps. The fact that the driving intensity and the
loss process are mutually dependent  means that our framework is
more general than the more standard doubly stochastic one which only
allows for a one-sided dependence.

The portfolio default intensity  is a derived process in our model.
It is shown to be a jump-diffusion that depends on the default
level. This dependence is governed by a set of multiplicative
factors - the so-called \emph{contagion factors}. These factors
enable a convenient semi-parametric specification which can be
calibrated to any set of consistent portfolio tranche quotes.
Furthermore, the fact that the model is two-dimensional and
Markovian means that
 efficient lattice implementations are available.

Our framework can be viewed as a low-dimensional short-rate version
of the approach described by Sch{\"o}nbucher in~\cite{Schon2005}. In
particular,  Sch{\"o}nbucher describes a Heath-Jarrow-Morton
(HJM)-like model of forward default transition rates which will in
general be non-Markovian and require Monte-Carlo simulations. The
BSLP model, on the other hand, imposes local drift conditions that
allow for fast lattice implementations, as indicated above.

Both models coincide in the local intensity limit where the
intensity process becomes a locally deterministic function of the
default level. In this case, we obtain a continuous-time,
one-dimensional  Markov chain driven by a deterministic Markov
generator. This can serve as a model in its own right, useful for
constructing an arbitrage-free interpolation of liquid index (e.g.\
\emph{iTraxx} or \emph{CDX}) tranche prices across strikes and
maturities. The interpolation can then be used  to price
non-standard index tranches consistently, which is often a problem
in the standard base correlation framework.

This is similar to the way local volatility models are used to price
exotic equity or FX  options off liquid  vanillas. While such a
framework can formally be viewed as a dynamic model, it is known
that local volatility models misspecify the {\it dynamics} of the
volatility process, and are therefore  ill-suited for pricing
instruments sensitive to the forward smile dynamics, such as forward
starting options and cliquets. This points to the need for
stochastic volatility models.

Similarly, what is needed in the present context, for the pricing of
multi-period credit derivatives is a stochastic {\it evolution} of
the loss distribution. This is obtained by making the Markov chain
generator stochastic. In this way, we arrive at the full-blown,
stochastic intensity version of the BSLP model.

Besides Sch{\"o}nbucher \cite{Schon2005}, our model is related to
a few other models suggested previously in the literature.
In particular, it
can be compared to the Markov ``infectious default'' model by Davis
and Lo \cite{DL}\footnote{which to our knowledge can be considered
the first dynamic model of the top-down type, without the ``down''
part. See also Frey and Backhaus \cite{Frey} for a related approach
to the portfolio dynamics with a mean-field interaction between
individual defaults.}. In their approach the portfolio default
intensity is piecewise deterministic, and follows a pure jump
process that jumps upon defaults. In our case we have a stochastic,
rather than piecewise deterministic, portfolio default intensity as
a result of introducing a diffusion component in the dynamics of the
driving factor. In addition, and perhaps more importantly, we set up
a semi-parametric framework that enables accurate and fast
calibration to market data. (See  Appendix C for a further
discussion of the relation between the two models.)

Our model can also be compared to the time-changed birth model of
portfolio loss dynamics of Ding et al.~\cite{Ding}. They consider a
{\it linear birth} process that has a self-affecting property
(controlled by a single parameter) and is therefore capable of
modeling credit contagion of credit losses in the portfolio. To have
more flexibility in the dynamics, an independent parametric
time-change process (CIR or quadratic Gaussian)  is introduced. This
plays a very similar role to the driving intensity in BSLP. The main
difference between the models is that BSLP is represented by a
\emph{non-linear death} process and has a semi-parametric
specification with a level-dependent amount of contagion controlled
by a set of contagion factors.
Our semi-parametric approach enables  accurate calibration to any
set of consistent quotes. However, this comes at the price of losing
the analytical tractability of Ding et al.\ and necessitating the
use of numerical (or approximate analytical) methods instead. An
additional difference between the two approaches is that, as
indicated above, we admit a back-impact of defaults on the driving
intensity process, which is absent in the model of Ding et al.

The paper is organized as follows. In Section~\ref{DTM} we present
the 1D local intensity version of the model.
Section~\ref{sec:dynamicBSLP} describes the 2D stochastic intensity
extension. We specify the default intensity process, forward
equations for the full 2D model, and its calibration. In
Section~\ref{sec:pricingAlgorithms} we summarize lattice-based
pricing algorithms for tranche, tranche options, forward-starting
tranches etc. Section~\ref{NumResults} contains detailed analysis of
numerical results, including calibration of the model and pricing of
index tranchlets and options on indices and tranches, as well as
analysis of conditional forward spreads implied by the model.
Furthermore, we explain how to calculate sensitivities in our
framework, and compute index deltas for tranches and tranche
options. Section~\ref{Conclusion} summarizes. In addition, four
appendices discuss  more technical model details. Appendix A derives
the relation between the next-to-default intensity and the portfolio
default intensity. Appendices B and C develop the continuous-time
formulation of BSLP under different assumptions on the stochastic
driver dynamics. Appendix D describes the adiabatic approximation to
BSLP  that can be used to construct a semi-analytical approach to
pricing  credit vanilla and exotic products.

\section{BSLP with Local Loss Intensity}
\label{DTM}

In this section we describe a local intensity version of the BSLP
model, where default intensities are assumed to be locally
deterministic (dependent on the loss level only). More specifically,
we construct a Markov portfolio default process whose marginal
default distributions will be consistent with any set of
arbitrage-free quotes for tranches on the portfolio. Our
construction is similar to the one-step Markov chain construction
proposed by Sch{\"o}nbucher \cite{Schon2005}.

The model presented here will form the basis of the full-blown BSLP model,
which will be introduced in Sect.~\ref{sec:dynamicBSLP}
as a stochastic generalisation of the framework developed here.

\subsection{Markov chain for portfolio default process}

We  construct a model for the default counting process\footnote{The
relation between $N_t$ and the portfolio loss $L_t$ is described in
 section~\ref{sec:DTMcalibration}. } $N_t$:
\beq
 N_t = \sum_{i=1}^N \mathbb{I}_{\tau_i < t}
\eeq
where $\tau_i$ is the default time of the $i^{th}$ name in the
portfolio, and $N$ is the initial number of portfolio names. The
indicator function is denoted by $\mathbb{I}$.

We assume that the counting process $ N_t $ is Markovian, and model
it as a continuous-time Markov chain with generator matrix $ A_t $.
We recall that off-diagonal elements of $ A_t = \{ a_{ij}(t) \}_{i,j
= 0}^{N}$ provide transition probabilities for infinitesimal time
intervals $ dt $:
\beq
 P[N_{t+dt} = j | N_t = i ] = a_{ij}(t) dt
\eeq for  $ j \neq i $.

Given the generators $ A_t $, the matrix of finite-time transition
probabilities $ P(N_T,T|N_t,t) $ with matrix elements
\[
 p_{ij}(t,T) = P[N_{T}=j ,T | N_t = i,t]
 \]
 satisfies the forward equation
 \beq
\label{forwardM} \frac{\partial P}{\partial T} = P A
\eeq
In what follows we restrict our analysis to a piecewise
time-homogeneous continuous-time Markov-chain, where different
generator matrices are applied for time intervals $ [0,t_1],
[t_1,t_2] $ etc.\footnote{Typically, $ t_1, t_2 , \ldots $ will be
chosen to be maturities of the tranches that we are calibrating to
(c.f. section~\ref{sec:DTMcalibration}).}. Note that for any
interval $[t_1,t_2]$ on which $ A_t$ is constant, $ A_t = A $, the
forward equation can be solved to give:
\beq p_{ij}(t_1,t_2) = \left( e^{(t_2-t_1) A  } \right)_{ij}
\eeq
Without the piecewise-homogeneity of $A_t$, we would need to use a
time-ordered exponential in the equation above, along with the
substitution $ (t_2-t_1) A  \rightarrow \int_{t_1}^{t_2} A_s ds $.

Clearly, default counting is a non-decreasing process, which means
that its generator should have zeros below the
diagonal\footnote{Equivalently, if instead of the number of defaults
$ N_t $ we model the number of surviving firms $ n_t = N - N_t $,
then the generator will have zeros above the diagonal.}.
Furthermore, assuming that at most one default can occur in the
infinitesimal time $ dt $, the generator can be taken to be
bi-diagonal. Since we are dealing with infinitesimal time-steps this
is a reasonable assumption, which makes a parsimonious description
of the model possible\footnote{Multiple instantaneous defaults have
been explored in \cite{Brigo}.}.

The simplest Markov non-increasing process with a bi-diagonal
generator is a well-known linear death process\footnote{Note that
usually death processes are formulated in terms of the decreasing
survival counter variable $ n_t = N - N_t $ rather than the default
counter variable $ N_t $.}. The constant generator $ A $ takes the
following form
\bea
\label{linearDeath} A =  \lambda \left(
\begin{array}{ccccccccc}
- N &  N & 0 & \cdots & 0 & 0  \\
0 & - (N-1)& (N-1)  & \cdots & 0 & 0  \\
\vdots \\
0 & 0 & 0 & \cdots & - 1& 1 \\
0 & 0 & 0 & \cdots & 0 & 0
\end{array} \right)
\eea
Here  $ \lambda $ is a parameter interpreted as the average
single-name default intensity of all names.

It is well know (see e.g. Feller \cite{Feller}) that the death
process gives rise to the binomial distribution for the number of
events (in our case, defaults):
\beq \label{binom} \pi(n,t) \equiv P[N_t=n] = \binom{N}{n} \left( 1
- e^{-\lambda t} \right)^n e^{- (N-n)\lambda t} \eeq
The binomial distribution (\ref{binom}) corresponds to the
case of non-interacting (``uncorrelated") credit exposures in the
portfolio, and is very far from distributions implied by market
prices of liquid tranches, for any choice of $ \lambda $.

To produce  a more interesting and non-trivial distribution of
defaults, we consider a minimal modification of the homogeneous
linear death process (\ref{linearDeath}). In particular, we make it
piecewise time-homogeneous to calibrate to tranche quotes at
different maturities. We also generalize it to a \emph{non-linear}
death process, where transition intensities are a general function
of the number of defaults $ n $ in the portfolio. We therefore
consider the following  generator:
\bea \label{nonlinearDeath} A_t = \lambda \left(
\begin{array}{ccccccc}
- f_{0}(t) N  & f_{0}(t) N & 0 & \cdots & 0 & 0  \\
0 & - f_1(t) (N-1) &   f_1(t) (N-1)  & \cdots & 0 & 0  \\
\vdots \\
0 & 0 & 0 & \cdots &   - f_{N-1}(t)&  f_{N-1}(t) \\
0 & 0 & 0 & \cdots & 0 & 0
\end{array} \right)
\eea

Clearly, the linear death process~(\ref{linearDeath}) is recovered
if all $ f_i(t) = 1 $. The generator (\ref{nonlinearDeath})
corresponds to the case where the next-to-default intensity in the
scenario where $ n $ names have defaulted is given by:
\beq
\lambda_{NtD}(n,t) = \lambda f_n(t)(N-n)
\eeq
If $ f_n(t) \neq 1 $, it means that the default intensity is a
non-trivial function of the default level and the resulting implied
default dynamics is contagious (self-affecting). Hence, in what
follows, the parameters $ f_i(t) $ introduced here will be referred
to as {\it contagion factors}.

As discussed in the next section, a suitable parametrization of the
contagion factors will enable calibration to any set of consistent
portfolio tranche prices.

The bi-diagonal generator provides a convenient low-dimensional
parametrization of the portfolio default process. Indeed, numerical
calculations of break-even tranche spreads according to equations
(\ref{DL}) and (\ref{PL}) (see below) require evaluation of loss
distributions at quarterly time steps. Probabilities of multiple
defaults within such periods cannot be neglected. Attempting to
directly parameterize these probabilities in a discrete Markov chain
framework would be a difficult task.

In contrast, the continuous-time framework allows the calculation of
all finite time, multiple default, probabilities by taking matrix
exponentials of the generator (\ref{nonlinearDeath}). This can be
done  efficiently if we neglect small differences in accrual
periods, as  described in the next section.

\subsection{Numerical implementation and calibration to tranche quotes}
\label{sec:DTMcalibration}

We assume that we are given  quotes for a set of tranches at
different maturities. Let $\vec{k}$  denote the ordered set of all
low strikes (including $100\%$ if we are calibrating to the index
level). For example, for iTraxx we have $\vec{k} = \{0\%, 3\%, 6\%,
9\%, 12\%,22\%,100\%\}$. Similarly, the set of quoted maturities is
denoted by $\vec{T}$, so that for example for iTraxx series 6 we
would have: $\vec{T} = \{$20-Dec-09, 20-Dec-11, 20-Dec-13,
20-Dec-16$\}$.

Since tranche payoffs are determined by the cumulative portfolio
loss, $ L_t $, and not the number of defaults, $N_t $, we need to
specify the relation between the two variables. The most general
assumption we can make within our model is that $L_t$ is a function
of $N_t$ only\footnote{In particular, this excludes time-dependent
recovery rates.}, i.e. $L_t = L(N_t)$. Typically, however, we can
set the recovery rate, $R$, to a constant value, in which case we
have: $L(N_t) = (1-R) N_t/N$, where $N$ is the initial number of
names in the portfolio.

In order to calibrate to the given set of quotes we need to choose a
parametrization of the contagion factors $f_i(t)$. To do this we
introduce a contagion function, $g(L_t,t)$, which is a function of
the portfolio loss level and time. This function is defined on the
interval $[0,1] \times [0,\infty)$ and takes values on the positive
real line: $[0, \infty)$. We define the contagion factors as:
\beq \label{contFacDef} f_i(t) = g(L(i),t) \eeq
where as above $L_t = L(N_t)$. Lastly, we restrict the functional
form of $g$ so that the number of free parameters exactly matches
the number of tranche quotes. To do this we assume that $g(l,t)$ is
piecewise constant in $t$ (for a given $l$) with node points given
by the quoted maturities, $\vec{T}$. In addition we take $g(l,t)$ to
be piecewise linear in $l$ (for a given $t$) between node points
$\vec{k}$ (and piecewise flat outside).

Using a multi-dimensional solver we can now calibrate the contagion
factors $f_k(t)$ to the observed tranche prices. This is done by
calculating all finite time loss distributions and using them along
with the standard pricing formulae (reviewed in the next section).
This results in  a calibrated generator of the default process:
\beq
\label{Lambda}
  a_{ij}(t) = \lambda (N-i) f_i(t) ( \delta_{i+1,j} - \delta_{i,j})\, \  \; i,j = 0,\ldots,N
\eeq
where $\delta_{i,j}$ denotes the Kronecker delta.
 As described previously, finite time transition probabilities
can be obtained from the generator (\ref{nonlinearDeath}) by matrix
exponentiation. This can be done using efficient numerical methods
such as the Pad{\'e} approximation\footnote{In practice, we can
neglect small differences in accrual periods, and use the Pad{\'e}
approximation to calculate the transition matrix $ P_1 $ for the
first period only. All other transition probability (two-period
etc.) matrices are obtained by repeated multiplication of $ P_1 $ by
itself.}.

\subsection{Pricing}
\label{sec:pricing}

 Once all finite-time loss transition probabilities are
calculated from the generator, the pricing of a tranche with lower
strike $ K_d $ and upper strike $ K_u $ (both defined as percentages
of the total portfolio notional $ N_0 $) with maturity $ T $ is done
using  standard formulae which we present here for the sake of
completeness.

In the following, we assume that the recovery rate $R$, is fixed for
all names in the portfolio. Then the default (or contingent or
protection) leg, $\DL$, of a tranche paid by the protection seller
to the protection buyer is given by
\beq \label{DL} \DL = N_0 \int_{0}^{T} B(0,t) dEL_t \simeq
N_0\sum_{i = 1}^{M} \frac{1}{2} \left( B_{i-1} + B_i \right) (
EL_{i} - EL_{i-1} ) \eeq
where the sum runs over all coupon dates $ T_i $, $ i = 1, \ldots, M
$, $ B_i \equiv B(0,T_i) $ is a risk-free discount factor, and
$EL_i$ is the tranche expected loss, given by:
\beq \label{EL} EL_i \equiv EL_{T_i} = \mathbb{E} \left[
L_{[K_d,K_u]}(T_i) \right] \eeq
%
%
%
where $L_{[K_d,K_u]}(t)$ is the tranche loss at time $t$ given by:
\beq
  L_{[K_d,K_u]}(t) =( L_{t} -
K_d)^{+}  - ( L_{t} - K_u)^{+}  \eeq

The premium leg (paid by the protection buyer to the protection
seller) is given by
\bea \label{PL} \PL(S) &=& S \cdot N_0 \sum_{i = 1}^{M} \Delta_i
\left( B_i \cdot EN_{T_i} - \int_{T_{i-1}}^{T_i}  \frac{ t -
T_{i-1}}{T_i - T_{i-1}}
B(0,t) dEN_t \right) \\
&\simeq& S \cdot N_0 \sum_{i = 1}^{M} \Delta_i B_i \frac{1}{2}
\left( EN_{i-1} + EN_i \right)    \nonumber \eea
where $ S $ is the tranche spread, $ \Delta_i $ is the day count
fraction and $EN_i$ is the expected tranche outstanding notional at
time $T_i$. It is defined by:
\beq \label{It} EN_i \equiv EN_{T_i} = \mathbb{E}[N_{[K_u,K_d]}(T_i)
] \equiv \mathbb{E} [ K_u - K_d - L_{[K_u,K_d]}]
 \eeq

The integral term in (\ref{PL}) represents the accrued coupon due to
defaults happening  between the coupon payments dates. The integral
is calculated using the standard  approximation: $ ( t -
T_{i-1})/(T_i - T_{i-1}) \rightarrow 1/2 $ and $ B(0,t) \rightarrow
B_i $.

The tranche par (or fair or break-even) spread,  $ S $, is
determined from the par equation $ \DL = \PL(S) $. Most index quotes
are given in terms of the par spread. For the equity tranche,
however, the market convention is to charge an upfront payment from
the protection buyer while fixing the running spread $S$ at 500 bp.

\subsection{Bootstrapping}
\label{Bootstrap}

It can be useful in practice to convert the calibration  into a
bootstrapping problem. This is possible if we use piecewise-flat
interpolation of the contagion function $g(l,t)$  in both time and
loss dimensions.

This can be easily seen by the fact that the calculation of the
par-spread, $S_{[K_d,K_u]}(T)$ (of a tranche with low strike $K_d$,
high strike $K_u$, and maturity $T$) will depend only on the
function $g(l,t)$ in the region $0 \leq t \leq T$ and $0 \leq l \leq
K_u$.

Because of the piecewise-flat assumption  we can calibrate the
contagion function values at the points $\vec{k}$ and $\vec{T}$
iteratively in the time and loss dimensions.

At this point we should note why we have not made the piecewise-flat
loss interpolation choice in the previous section. This is because
for typical tranche quotes such as iTraxx, the piecewise-flat
assumption is not very good for the equity region and  results in
poor calibration. This is essentially due to the thickness of the
standard tranches, e.g.\ $3\%$ for junior iTraxx and CDX tranches.
Indeed, if we calibrate to thinner tranches then the piecewise-flat
assumption will become more viable.

In particular, in the limit where we consider tranches that are
exactly one default wide, piecewise-linear and piecewise-constant
interpolations become equivalent. This means that bootstrapping will
produce very accurate calibrations provided the input quotes are
arbitrage free.

\subsection{Discussion}

In this section we discuss the applicability of the local intensity model presented above.

First we note that the price of a tranche depends only on the
marginal default distributions at a set of time points but not on
the inter-temporal correlations between defaults at different times.
In other words, tranches are single-period instruments.

Let us hence assume that we have found a set of marginal default
distributions such that prices of all liquid tranches are matched.
It is well known that this is {\it not} equivalent to, and not by
itself sufficient for, defining a dynamic model.

This is best seen by considering the
 {\it path probability} for  the default counting process $ N_t $ to pass through
levels $ [N_h, N_{2h}, \ldots , N_{nh}] $ at times $ t = [h,2h,
\ldots, nh] $. Using the chain rule, we write
\beq \label{condProb}
P \left[ N_h, N_{2h}, \ldots , N_{nh} \right] = P \left[ N_h \right]
P\left[N_{2h} | N_h \right]  \ldots P \left[ N_{nh} | N_{h} N_{2h}
\ldots N_{(n-1)h} \right]
\eeq
The important point here is that tranche market prices do {\it not}
constrain the conditional transition probabilities
in~(\ref{condProb}). The only thing that can be uniquely determined
from the market\footnote{Assuming its completeness, i.e.\ the
availability of tranche price quotes for all strikes and
maturities.} are marginal loss distributions at various times
(implied distributions) and {\it one-step} transition
probabilities\footnote{This was demonstrated by Britten-Jones and
Neuberger (BJN) \cite{BJN} in the context of stochastic volatility
models for equity options. Modifications of their derivation for the
present context of tranche pricing are straightforward.} $ P[N_{t+h}
| N_t] $.

 The calculation of the path probability (\ref{condProb}) reduces
to a product of one-step transition probabilities only in the
special case where $ N_t $ is a Markov process. However, whenever
additional stochastic risk factors are present in the model this
would generally not be the case\footnote{Let $ M_t = (N_t, \ldots )$
be the vector combining all relevant state variables. If the
multi-dimensional process $ M_t $ is jointly Markov, then its
individual components $ N_t \, \ldots $ are generally non-Markov
when viewed separately.}. Hence it is only by making further
assumptions that a set of marginal distributions for different time
horizons, originally viewed as a disjoint set, can be ``glued''
together within a unique {\it dynamic} process.

In the local intensity setting, we assumed that the portfolio
default counting variable, $ N_t $, is the only relevant state
variable, and its dynamics was further supposed to be Markov. This
assumption allowed us to construct a model which can calibrate
perfectly to any set of tranche prices.

Formally, such a 1D Markov chain model may be viewed as a dynamic,
multi-period, model since the Markov property allows one to
calculate all default transition probabilities. However, the
assumptions underlying this model are likely to be too extreme. In
particular, defaults are rare events and we expect portfolio spreads
to move stochastically even in the absence of defaults, which cannot
be achieved in the present 1D setting.

Even if we are willing to ignore the requirement of any realism in
modeling portfolio spreads, we face the problem that the
default-only model is completely specified once calibrated. It is
likely that we will need extra flexibility to adjust the dynamic
properties of the model when pricing exotic derivatives.

For example, stochastic loss intensities are needed if we want to
have control over the tranche option implied volatilities. Another
issue is that the 1D model implies high inter-temporal correlations
between losses at different time horizons. Again we need to introduce
stochastic intensities to gain more control.

Furthermore,  experience with equity and FX modeling suggests that
local volatility-type models tend to give rise to strong
time-inhomogeneity of dynamics which, while allowing for a good fit
to today's option prices, makes the evolution of option prices (or
equivalently, implied volatility) unrealistic.

For these reasons the 1D model is unlikely to be adequate for
pricing path-dependent instruments or/and derivatives that are
sensitive to the {\it dynamics} of the loss process. Note that most
 exotic credit derivatives (tranche options, forward-starting
tranches, leveraged super-seniors etc.) are of this type. To model
these instruments, we need a dynamic extension of the framework
developed so far. This will be presented in the next section.

To summarize, the local intensity version of our model appears
inadequate for pricing credit exotics, however it is useful on two
other counts:
%
\begin{itemize}
\item It provides a very convenient way to parameterize a set of
observable liquid tranche quotes. This can be used for an intuitive
interpolation-based pricing of non-standard tranches off the liquid
ones.
\item It can serve as a first step towards the second, dynamic
stage of the model.
\end{itemize}

\section{Dynamic Intensity BSLP model}
\label{sec:dynamicBSLP}

\subsection{Stochastic loss intensity}

We construct a dynamic version of the model presented above by
promoting the constant $ \lambda $ in the generator
(\ref{nonlinearDeath}) to a stochastic driving process $ Y_t $. This
means that the next-to-default intensity $\lambda_{NtD}(N_t,Y_t,t)$
is now stochastic and is given by:
\beq \label{ntD} \lambda_{NtD}( N_t = k, Y_t, t)  \equiv (N- k)
F(k,t)Y_t
 \eeq
Here we have introduced the contagion factors $F(k,t)$, which, as
before, are a deterministic function of default level $k$ and time
$t$. Note that these factors will, in general, differ numerically
from the factors $f$ defined in equation~(\ref{Lambda}). We will
occasionally refer to the former and the latter as the 2D and 1D
contagion factors, respectively.

We consider the general form of the SDE of the driving process
$Y_t$: \beq \label{SDEY} dY_t = \mu(Y_t,t) dt + \sigma(Y_t,t) dW_t +
\gamma(Y_t,t) dN_t \eeq Several  specifications of the dynamics of $
Y_t $ are possible. We have considered a mean-reverting, log-normal
process with
\bea
\label{logNormal}
\mu_{ln}(Y_{t-},t) &=& a Y_{t-}
( \ln \theta_t - \ln Y_{t-}) \\ \nonumber
 \sigma_{ln}(Y_{t-},t) &=&  \sigma Y_{t-} \\
 \gamma_{ln}(Y_{t-},t) &=&  \gamma Y_{t-} \nonumber
 \eea
and a mapped OU form $ Y_t = q(y_t) $ where $ q(x) $ is a function
ensuring positivity of the mapping and $ y_t $ is an
Ornstein-Uhlenbeck (OU) process with
\bea
\label{OUy}
\mu_{OU}(y_{t-},t) &=& a ( \theta_t - y_{t-})  \nonumber \\
 \sigma_{OU}(y_{t-},t) &=& \sigma \\
 \gamma_{OU}(y_{t-},t) &=& \gamma \nonumber
 \eea

It will be important later that the $Y_t$ process does not depend on
the default level $N_t$. It can, however, include a jump component,
$dN_t$, which depends on \emph{changes} in the default
level\footnote{Note that a slightly more sophisticated form of the
jump term in (\ref{SDEY}) $ \gamma_{ln}(Y_t,t) = \gamma p_t Y_t $ or
$ \gamma_{OU}(Y_t,t) = \gamma p_t $ (where $ p_t $ is a Bernoulli
random variable) could be considered as well. We do not pursue this
modification in what follows.}.

Note that for both  (\ref{logNormal}) and (\ref{OUy}) the model
obtained in the limit $ \sigma \rightarrow 0 $, $ a \rightarrow
\infty $ and $ \gamma \rightarrow 0 $  becomes equivalent to the
local intensity model described in the previous section.
Alternatively, in the zero volatility limit $ \sigma \rightarrow 0 $
with $ \gamma \neq 0 $, we obtain a pure jump process for $ Y_t $
driven by jumps in $ N_t $, which fits into the
piecewise-deterministic Markov framework of Davis and Lo (DL)
\cite{DL}\footnote{The difference being the different number of
possible states for $ Y_t $ - in the original DL model, the loss
intensity can only assume two values, while here it would be
continuous until of course we discretize it for practical
implementation.}.

\subsection{Index swap intensity process in BSLP}

We now consider in more detail the BSLP dynamics implied by our
definition of the next-to-default intensity~(\ref{ntD}). This is
done most conveniently by considering the SDE for the index swap
intensity\footnote{This is the stochastic intensity that prices the
index correctly, assuming it is a single name CDS. For a homogeneous
portfolio this is equal to the individual single-name portfolio
intensities.}
 $ \lambda_s (t) $. The reason for focussing on the swap
intensity is that it  represents the average portfolio single name
intensity. In the absence of contagion this can be expected to be
largely independent of the portfolio size. This is not the case for
the next-to-default intensity, which even in the independence case
is proportional to the remaining number of names in the portfolio.

In what follows we assume that recovery  rates are constant and
homogeneous across the portfolio. If in addition we assume that
defaults cannot happen simultaneously,  then $\lambda_s$ is obtained
from the first-to-default intensity by division by the number of
surviving names\footnote{ Note that the relation (\ref{lambdaS})
expressing the portfolio loss intensity  $ \lambda_s(t) $ as a
product of the driving factor $ Y_t $ and the contagion factors $
F(N_k,t) $ resembles the relation between the stochastic time change
rate $ \nu_t $ and the generator of the initial Markov chain for the
default process in the model by Ding et al \cite{Ding}, and can be
interpreted correspondingly. }:
\beq \label{lambdaS} \lambda_s (t) = \frac{ \lambda_{NtD}(N_t, Y_t,
t)}{N - N_t} =   Y_t F(N_t,t) \eeq
While the relation between
$\lambda_s$ and $\lambda_{NtD}$ may  look intuitively obvious, we
present a formal derivation in Appendix A.

To derive the SDE we first note that using It\^{o}'s lemma for a
pure jump process $ N_t $, we obtain
 \bea
 \label{Ito}
dF(N_t,t) &=& \frac{\partial F_t (N_{t-},t)}{\partial t} dt  + (F(N_t,t) - F(N_{t-},t)) d N_t \\
 &\equiv& F'(N_{t-},t) dt + \Delta F(N_t,t) d N_t \nonumber
 \eea
where $N_{t-} = N_t-1$ if there is a default at time $t$.
 Now the dynamics of $ \lambda_s(t) $ are determined as follows:
\beq \label{dynlambdaS}
d \lambda_s(t) = d( Y_t F(N_t,t)) =
F(N_{t-},t) dY_t + Y_{t-} d F(N_t,t) + \gamma(Y_{t-},t)\Delta F(N_t,t)dN_t
\eeq

Then (\ref{dynlambdaS}) together with (\ref{Ito}) and (\ref{SDEY})
yield the following SDE for the swap intensity $ \lambda_s(t) $:
\beq \label{SDElambdas} d \lambda_s(t) = \mu_s(\lambda_s,N_t,t) dt +
\sigma_s(\lambda_s,N_t,t) dW_t + \gamma_s(\lambda_s,N_t,t) dN_t \eeq
where \bea \label{coeffs} \mu_s(\lambda_s,N_t,t) &=& F(N_{t-},t) \mu
\left( \frac{\lambda_s}{F(N_{t-},t)},t \right) +
\lambda_s \frac{ F'(N_{t-},t)}{F(N_{t-},t)} \nonumber \\
\sigma_s(\lambda_s,N_t,t)  &=& F(N_{t-},t) \sigma \left( \frac{\lambda_s}{F(N_{t-},t)},t \right) \\
\gamma_s(\lambda_s,N_t,t) &=& \left( F(N_{t-},t)+\Delta
F(N_t,t)\right) \gamma \left( \frac{\lambda_s}{F(N_{t-},t)},t
\right) + \lambda_s \frac{ \Delta F(N_t,t)}{F(N_{t-} ,t)} \nonumber
\eea Equations (\ref{coeffs}) determine the $N_t $-dependent (and
generally non-linear) transformation needed to obtain the
coefficient functions of the SDE (\ref{SDElambdas}) from the
coefficient functions of the SDE (\ref{SDEY}). It is interesting to
consider these relations for the example of the log-normal mean
reverting (Black-Derman-Toy or BDT) specification (\ref{logNormal}).
For this case, we obtain
\bea \mu_s(\lambda_s,N_t,t)/\lambda_s(t-) &=& a \left( \ln
(F(N_{t-},t) \theta_t)) - \ln \lambda_s(t-)\right)
+  \frac{ F'(N_{t-},t)}{F(N_{t-},t)} \nonumber \\
\sigma_s( \lambda_s,N_t,t) / \lambda_s(t-) &=& \sigma   \\
\gamma_s( \lambda_s, N_t,t) / \lambda_s(t-)  &=&  \gamma  +  \frac{
\Delta F(N_t,t)}{F(N_{t-} ,t)}(1+\gamma) \nonumber
\eea
This is an interesting result. It shows that if we start with a
jump-diffusion BDT specification for the driving $Y_t$-process, then
the SDE for the quasi-observable swap intensity process  $
\lambda_s(t) $ is obtained from that of $ Y_t $ by adjusting its
drift and jump intensity, while keeping the volatility constant.

We can also clearly see  that there are two qualitatively different
types of contagion implied by the model. The coefficient $\gamma_s(
\lambda_s, N_t,t)$ determines the jump size of the swap intensity
contingent on portfolio defaults. The jumps in intensity will
mean-revert over time, hence we say that $\gamma$ is responsible for
\emph{transient contagion}. Note  that even if we start with the
pure diffusion case $ \gamma = 0 $ for the process $ Y_t $, the
intensity $ \lambda_s(t) $ acquires a jump component with the  jump
size determined  by the relative change of the 2D contagion factor
$F(N_t,t) $.

The second important effect of portfolio defaults is that they
change the mean reversion level of the swap process. This is
referred to as \emph{permanent contagion} and is driven  by the
contagion factors $F(N_t,t)$.

This can be compared with the formalism  suggested by Lopatin and
Misirpashaev (LM) \cite{LM}. In their formulation, they start with a
stochastic next-to-default intensity $ \lambda^{LM}(N_t,t) $ where
the $N_t $-dependence arises entirely through the drift term.
Furthermore, LM do not admit jumps in the SDE for $
\lambda^{LM}(N_t,t) $, in contrast to the BSLP formulation. Hence,
in the LM model next-to-default intensities do not jump upon
defaults but gradually diffuse to a new mean-reversion level.

However, jumps are present in the SDE for the swap intensity, which
is related to the next-to-default intensity by the general formula
(\ref{lambdaS}):
\beq
\label{lambdaSCM}
\lambda_s^{LM}(t) = \frac{\lambda^{LM}(N_t,t)}{ N - N_t}
\eeq
Hence, we have:
\beq \label{SDE2LM} d \lambda_s^{LM}(t) = \frac{1}{N-N_t} d
\lambda^{LM}(N_t,t) + \lambda^{LM}(N_t,t) d \frac{1}{N-N_t} \eeq
To evaluate the second term, we use It\^{o}'s lemma for jump processes,
assuming in addition that $ d N_t \ll N - N_t $ so that we can use an approximation:
\beq
\label{Ito2}
d  \frac{1}{N - N_t} = \left( \frac{1}{N - N_t} -
\frac{1}{N - N_{t-}} \right) dN_t \simeq \left( \frac{1}{N - N_t}
\right)^2 dN_t
\eeq
%

This means that  in the LM model jumps in the portfolio swap
intensity $ \lambda_s(t)$ are solely due to the reduction of the
portfolio as defaults occur. In addition the jump size scales down
in a fixed manner as $ (N- N_t)^{-2} $. This is in contrast to the
BSLP model where the jump size of the swap intensity depends on the
amount of contagion in the data, expressed via the 2D contagion
factor $ F(N_t,t)$.

\subsection{ Forward equation and transition probabilities in 2D}

We postulate that the 2D dynamics of the pair $(N_t,Y_t)$ in BSLP
model is Markovian. We find it convenient to use a simple
discrete-time formulation of the model in what follows, with a time
step $ dt $ considered as small but finite. A continuous-time
formulation of BSLP may still be of interest from both theoretical
and practical viewpoints. This is discussed in some detail in
appendices B to D.

Discretizing the range of $ Y_t $ to a finite set $ \{ Y_ n \}$, the
system is described in terms of the joint marginal probabilities
\beq \label{joint} \pi(j,n,t) \equiv P \left[ N_t = j, Y_t = Y_n
\right]
\eeq
and conditional transition probabilities
\beq \label{trans} p_{jm| kn}(t, t + dt) \equiv P \left[ N_{t + d t}
= k, Y_{t + dt} = Y_n | N_{t} = j, Y_{t} = Y_m \right] \eeq
The forward equation takes the form
\beq
\label{forward}
\pi(k,n,t + dt) = \sum_{j,m} p_{jm| kn}(t, t + d t) \pi(j,m,t)
\eeq
We note that due to the composition law of probabilities we have the following relation
\bea
\label{composition}
& &  p_{jm| kn}(t, t + dt)  \\
& & =  P \left[ N_{t +  dt} = k | N_{t} = j, Y_{t} = Y_m \right] \,
P \left[  Y_{t + dt} = Y_n |  Y_{t} = Y_m,  N_{t} = j,
 N_{t + dt} = k \right] \nonumber
\eea
The form of the second factor
\beq
\label{YtransMat}
P_{mn}(j,k,t) \equiv  P \left[  Y_{t + dt} = Y_n |  Y_{t} = Y_m,  N_{t} = j,
 N_{t + dt} = k \right]
 \eeq
is fixed by the SDE (\ref{SDEY}). Note that here we used the fact
that our SDE for $Y_t$ does not depend on the default level $N_t$
but only on the change in default levels $dN_t$.

The formulae presented so far are completely general. BSLP is
defined by the  functional form of the first factor
in~(\ref{composition}).
\beq
\label{condLossTransProb} g_{jk}(Y_m,t) \equiv  P \left[ N_{t + d t}
= k  | N_t = j, Y_t = Y_m \right]
\eeq
This is determined by our
definition of the next-to-default intensity in equation~(\ref{ntD})
as follows:
\bea \label{ansatz} g_{jk}(Y_m,t) = \left\{  \begin{array}{lll} dt
(N-N_t) Y_m F_j(t) + O(dt^2)
& \mbox{if $ k = j+1 $} \\
1 -  dt (N- N_t)Y_m F_j(t) + O(dt^2) & \mbox{if $ k = j $} \\
O(dt^2)  & \mbox{otherwise}
\end{array}
\right.
\eea
where we have introduced $F_j(t) \equiv F(N_t=j,t)$.

\subsection{Tree discretisation}

To turn the above into a practical scheme, we discretise BSLP on a tree. To do this we introduce a
discrete timeline $t_0,t_1, \ldots, t_n$ with (finite) time step $\Delta t = t_{i+1} - t_i$.

We assume that $Y_t$ and $F_k(t)$ are piecewise constant between timeline points.

The transition probabilities defined in equation~(\ref{ansatz}) are given to first order in $dt$. If
we wanted to use these equations directly this would require very small (daily) time steps in the tree.

In practice our discretisation step $\Delta t$ is likely to be
larger. In this case we can no longer assume that there is only one
default over the time period. Our conditional transition
probabilities are now given by:
\beq \label{largeStepTransProb}
P[N_{t+\Delta t} = k | N_t = j, Y_t = Y_m] = \left( e^{\Delta t Y_m
\hat{A}_t} \right)_{jk}
\eeq
where $ \hat{A}_t $ is defined as in (\ref{nonlinearDeath}) with the
substitution $ \lambda \rightarrow 1 $ and $ f_i \rightarrow F_i(t)
$.
This expression generalizes (\ref{ansatz}) to the case of a finite
time step $ \Delta t $. Note that for a small but finite $ \Delta t
$, (\ref{largeStepTransProb}) coincides, to the first order in $
\Delta t $, with (\ref{ansatz}) integrated over the interval $ [t,
\, t + \Delta t] $ assuming that $ Y_t $ and $ F_k(t)$ do not change
in this interval. In other words, (\ref{largeStepTransProb}) can be
interpreted as the transition probability matrix in a conditional
discrete-time Markov chain obtained by ``freezing'' the random
variables $ Y_s $ and $ F_k(s) $ for all $ t \leq s \leq t + \Delta
t $ at their values at time $ t $.

Given equation~(\ref{largeStepTransProb}) we can construct a discretisation of BSLP on a 2D tree. In
particular, the factorisation in equation~(\ref{composition}) means that to calculate 2D transition
probabilities we can proceed in two steps:
\begin{enumerate}
\item Discretise the $Y_t$ process on a lattice or tree and calculate all 1D $Y$-transition probabilities.
\item Calculate the conditional transition probabilities in~(\ref{largeStepTransProb}). As mentioned
previously, efficient numerical methods, such as the Pad\'{e}
expansion, for calculating matrix exponentials are available.
\end{enumerate}
The probabilities $p_{jm| kn}(t, t + dt)$ are then given by the
product of the probabilities calculated in the two steps, as
in~(\ref{composition}).

\subsection{Two dimensional calibration}
\label{calibration}

BSLP is specified by the choice of contagion factors, $F(N_t,t)$ and
the parameters governing the $Y_t$-process~(\ref{SDEY}). As such
there are two possible routes for calibrating the model to external
data.

We could fix the form of the 2D contagion factors $ F(N_t,t) $, and
try to fit the $ Y_t $ to liquid tranche prices.
Alternatively, we could postulate some fixed law for $ Y_t $ and
calibrate the model in terms of the $ F(N_t,t)$. In this paper, we
explore the second route\footnote{Note that Lopatin and Misirpashaev
\cite{LM} essentially choose the first route by making their analog
of the $Y_t$-process explicitly dependent on  the $ N_t $-variable,
while effectively taking the factors $ F(N_t,t) $ to unity.
Computationally, the two  routes turn out to be nearly identical,
however as we have seen the difference in parametrization does
matter as it leads to observable effects in the dynamics of the swap
intensity. }.

In more detail, once we have fixed the parameters that govern the
$Y_t$-process,  we can calibrate the contagion factors $F(N_t,t)$
exactly as we do in Sect.~\ref{DTM}. In particular, we can still use
the bootstrapping algorithm as described. The only difference is
that now tranche prices are calculated on a 2D tree. This makes
calibration slower, although the speed is still acceptable (about
1-2 min for a standard index portfolio with calibration to four
standard maturities).

However, a much faster calibration algorithm is also available.
This is described in the next sections.

\subsection{Fast calibration by forward induction}
\label{BJN}

In this section we present a fast calibration algorithm that enables
a re-calibration of the full 2D model starting from a calibrated 1D
default tree (lattice) constructed  in section~\ref{DTM}. It uses a
recursive procedure of ``integrating in" the stochastic loss
intensity. Our method is similar to the algorithm developed by
Britten-Jones and Neuberger (BJN) in the context of stochastic
volatility modeling. In  turn, the BJN method is closely related to
the Markov projection method used by Lopatin and Misirpashaev
\cite{LM}\footnote{When the BSLP model was developed, we were not
aware of V. Piterbarg's paper on the Markov projection method (V.
Piterbarg, ``Markovian projection method for volatility
calibration'', available at http://ssrn.com/abstract = 906473), and
instead referred to this approach as the BJN method. It appears
that, at least for the setting used in this paper, the two methods
are very similar, if not identical. Note that while Piterbarg does
not cite BJN, both BJN and Piterbarg cite the work by Dupire on the
link between stochastic and local volatility models. Dupire's
approach seems to provide a common basis for both the BJN and Markov
projection methods.}.

We start by summing over $n$ in the forward equation
(\ref{forward}), to get the probability unconditional on $Y_{t+dt}$:

 \beq \label{reducedForward} \pi(k, t + dt) = \sum_{j,m} P \left[
N_{t+dt} = k| N_t = j, Y_t = Y_m \right] \pi(j,m,t) \eeq
This is the marginal $N$-probability at time $ t + dt $. Assuming
that the model is constructed as described above (i.e. in two
stages) we have the 1D $ N_t $-tree (lattice) calibrated to the
observed set of tranche quotes. Marginal probabilities in the full
2D model should match ones calculated in the 1D model. Moreover, we
can relate the marginal probability at $ t + d t $ with the one at $
t $ using the forward equation in the 1D model:
\bea \label{1Dforward} \pi(k, t + d t) &=&
 a_{k-1, k}  \pi(k-1, t)dt
+  (1 -  a_{k, k}dt) \pi(k, t) + O (d t^2)  \eea
%
%
%
We now re-write equations~(\ref{ansatz}) in the following form:
\bea
\label{ansatzBJN} P \left[ N_{t+dt} = k | N_t = j, Y_t = Y_m \right]
= \left\{  \begin{array}{lll} dt (N-j) Y_m q_j(t)f_j(t) + O(dt^2)
& \mbox{if $ k = j+1 $} \\
1 -  dt (N- j)Y_m q_j(t)f_j(t) + O(dt^2) & \mbox{if $ k = j $} \\
O(dt^2)  & \mbox{otherwise}
\end{array}
\right. \eea
where $f_j(t) \equiv f(N_t=j,t)$ are the contagion factors
introduced in equation~(\ref{Lambda}), and $q_j(t) \equiv
q(N_t=j,t)$ are the \emph{drift adjustment factors}. We assume that
the contagion factors $f_j(t)$ are chosen such that the 1D
$N_t$-tree is calibrated to the market data. Note that as long as
the factors $q_j(t)$ are not yet defined, the parametrization
in~(\ref{ansatzBJN}) is only a matter of convenience, and is
completely equivalent to (\ref{ansatz}).

 By substituting (\ref{ansatzBJN}) into~(\ref{reducedForward}) and~(\ref{Lambda})
 into~(\ref{1Dforward}) and by comparing the two
resulting
  forward equations,
  we obtain the following constraint on the drift
 adjustment factors $ q_j(t) $:
 \beq
 \label{driftConstraint}
  q_{N_t}(t) f_{N_t}(t) \sum_{m} Y_m \pi(N_t,m,t) =  \lambda f_{N_t}(t) \pi(N_t,t)
\; \Leftrightarrow \; q_{N_t}(t) = \frac{\lambda}{ \mathbb{E} \left[Y_t|N_t \right]}
 \eeq
 The no-arbitrage drift constraint just derived is our short-rate analog of the no-arbitrage
 drift constraints in the HJM-like construction of Sch{\"o}nbucher \cite{Schon2005}, where the
 drift corrections are typically non-local in time. In contrast, our drift constraints (\ref{driftConstraint})
 are local in both $ N_t $ and $ t $, and are thus amenable to a fast lattice implementation.
  Note that our drift constraint coincides with the $ d t \rightarrow 0 $ limit of the drift constraint
  obtained by BJN \cite{BJN} using a more involved
  argument (and in a different context). Also note that the same drift condition
  (albeit with a different notation) is obtained by LM \cite{LM} who further identify it
  as a special case of a general Markovian projection approach.

 We now use the BSLP drift constraint (\ref{driftConstraint}) in order to set up a convenient and fast calibration method for the 2D BSLP tree
 based on a combination of the 1D calibration of the $N_t$-tree and a forward induction method.

 We assume that the 1D calibration of the $N_t $-tree is performed as discussed above. We start with
 the initial conditions for the 2D and 1D probability distributions, correspondingly,
 \beq
 \label{initCond}
 \pi(N_0,Y_m, 0) = \delta_{N_0,0} \delta_{m,\hat{m}} \; , \; \pi(N_0,0) = \delta_{N_0,0}
 \eeq
 where $ \hat{m} $ is the index corresponding to the initial value $ Y_0 $ (which we assume to be
 known). Using Eq.(\ref{driftConstraint}), we solve for $ q_{N_0}(0) $:
 \beq
 \label{firstQ}
 q_{N_0}(0) = \frac{ \lambda \pi(N_0,0)}{
 \sum_{m} Y_m \pi(N_0, Y_m,0)} = \frac{\lambda}{Y_{\hat{m}}}  \; , \; N_0 = 0
 \eeq
 Note that for $ N_0 \neq 0 $ the correction factors $ q_{N_0}(0) $ are undefined. However, this does not
 pose any problem as such states are unachievable at time $ t = 0 $, and therefore play no
 role in the dynamics whatsoever. If desired, these parameters can be assigned
 some definite dummy
 values that would not have any impact on the numerical results reported below.
 We now use the forward equation
 \bea
 \label{forward2}
 \pi(k,n,t+ dt) &=& \sum_{m} g_{k-1,k}(Y_m,t) P_{mn}(k-1,k,t) \pi(j,m,t)\\
 &+&
 \sum_{m} g_{k,k} (Y_m,t) P_{mn}(k,k,t) \pi(k,m,t) \nonumber
\eea
(where the conditional $ Y_t$-transition matrix $ P_{mn}(i,j,t) $ is
defined in (\ref{YtransMat})) at $ t = 0 $ to calculate the joint
probability distribution $ \pi(k,n, dt) $. Then we calculate the
drift adjustments $ q_j(dt) $ for the second period from the drift
constraint  (\ref{driftConstraint}). Next we use it to calculate $
\pi(k,n, 2dt) $ from the forward equation (\ref{forward2}) evaluated
at $ t = dt $, and so on. As a result, we have a full 2D BSLP tree
calibrated to the set of tranche quotes using a fast and effective
algorithm.

To avoid possible misunderstanding, a comment  on the meaning of the
above procedure is in order here. The drift condition
(\ref{driftConstraint}) provides a way to specify the 2D dynamics
once the 1D dynamics of $ N_t $ is fixed. Is this equivalent to
saying that the 2D dynamics is uniquely (once the SDE for $ Y_t $ is
specified) fixed by observed tranche prices? The answer is no, of
course, as market incompleteness means there is no unique
correspondence between tranche prices and marginal default
distributions. The meaning of the above procedure is rather to {\it
pick} a 1D model that matches all tranche quotes, and then calibrate
the full 2D dynamics to this {\it model}, not to the data directly.

\subsubsection{Fast calibration in practice}

 The algorithm of forward induction-based calibration of the BSLP model just presented is simple
 and intuitive, however it is not ideal from a practical viewpoint, as it assumes that
the time steps are small enough to justify the use of a binomial
approximation for the one-step default process. In practice, this
means we should take daily steps, which
may slow down calibration and pricing. If we want to be able to have a tree with
larger steps (e.g. monthly), we need to use a different method.


Several alternatives of different complexity can be considered at this point.
The first one is to generalize the relation (\ref{ansatzBJN}) to the case of non-infinitesimal time
steps $ \Delta t $. We have considered the following specification
\bea
\label{ansatzBigStep}
P \left[ N_{t+\Delta t} = k | N_t = j, Y_t = Y_m \right] = \left\{  \begin{array}{lll}
Q_{jk} (t) \left( e^{ \Delta t Y_m A_t} \right)_{jk}
& \mbox{if $ k > j $} \\
1 -  \sum_{k \neq j}  P \left[ N_{t+\Delta t} = k | N_t = j, Y_t = Y_m \right] & \mbox{if $ k = j $} \\
0  & \mbox{otherwise}
\end{array}
\right. \eea which is similar to parametrizations used for
stochastic volatility models by BJN \cite{BJN}. Here $ A_t $ stands
for the calibrated generator of the 1D default-only model, and $
Q_{jk}(t) $ are drift adjustment factors similar to the factors $
q_j(t) $ in (\ref{ansatz}). We may interpret the term $ e^{ \Delta t
Y_m A_t} $ in (\ref{ansatzBigStep}) as an ``initial guess'' or a
``prior'' for the finite-time conditional transition probability,
which is then corrected by a set of multiplicative drift adjustment
factors $ Q_{jk}$ using a finite-time version of
(\ref{driftConstraint}). Note that  the ansatz (\ref{ansatzBigStep})
is somewhat non-symmetric with respect to its dependence on $Y_m $,
i.e. the drift adjustments $ Q_{jk} $ with $ j \neq k $ are assumed
to be independent of $ Y_m $, while the diagonal adjustment $ Q_{jj}
$ is implicitly dependent on $ Y_m $, as required by conservation of
probability.

We have implemented this scheme and tested it on a number of
portfolios. Unfortunately, we have found that while this method
works well for some portfolios, it develops numerical instabilities
for others. However, we were able to find a simple practical
solution to this problem, which involves using the adjustment given
in the first line of (\ref{ansatzBigStep}) for both off-diagonal and
diagonal transitions, and then rescaling all probabilities by a
common factor chosen to ensure the correct normalization. Note that
this produces a more uniform dependence on $ Y_m $ than in the
ansatz (\ref{ansatzBigStep}). This algorithm was found to be stable
and accurate in all cases we tested\footnote{ The price we have to
pay with this method is that it introduces a small mismatch between
tranche prices evaluated in the 2D and 1D versions of the model, but
mismatches were found to be negligibly small for all test portfolios
we considered.}, with model outputs similar to those obtained with a
direct 2D calibration described above.

Lastly, we note that in addition to the numerical methods discussed
so far, analytical approximations to the model are possible as well.
In particular, we can consider the adiabatic approximation, which is
expected to produce accurate results as long as the characteristic
time of changes in spreads is much smaller than those in the loss
counting variable. This assumption is expected to hold in reality
(spreads change daily, while defaults are rare events). Derivation
of the adiabatic approximation is presented in Appendix C.
Interestingly, this approach produces an analytical formula for the
drift adjustments $ q_j(t) $  similar to the one defined by
(\ref{driftConstraint}).

\section{Pricing algorithms}
\label{sec:pricingAlgorithms}

Models discretised on a tree or lattice are particularly suitable
for pricing products which have a payoff that is amenable to a
backward recursion algorithm. Typically, these are products with
embedded optionality, such as  tranche options or leveraged
super-seniors.

In addition, tree pricing\footnote{It is also possible to construct
a Monte Carlo implementation of BSLP to price path dependent
products. This is not explored in this paper.}
 of products that are
weakly path-dependent is also tractable. By weakly path-dependent we
mean that the payoff depends on the loss path at only a few points
in the past. A forward tranche is an example of such a product.

In this section we take a closer look at how tree pricing algorithms
can be applied to the products mentioned above. Note that the
algorithms presented are not specific to the BSLP model. In the
following we also assume that defaults and losses are related simply
by $L_t = (1-R)N_t$ and hence we will only talk about losses below.

\subsection{Tranche pricing by backward induction}

Let $ \DL(t) $ and $ \PL(t) $ be the default and the premium legs at
time $ t $ of a tranche with strikes $ K_d $ and $ K_{u} $ with
maturity $ T $. Let  $ 0 = t_1, \ldots, t_M = T$ be the time grid
and $ T_n $ ($ n = 1,2,\ldots,M_c $) be the coupon payment dates on
the grid. Then, neglecting the accrued coupon, we obtain the
following expressions:
\bea \label{DLPL} \DL(t_i) &=& \mathbb{E}_i \left[ \sum_{j=i}^{M-1}
B(t_i,t_{j+1})
\left( L_{[K_d,K_u]}(t_{j+1}) - L_{[K_d,K_u]}(t_j) \right) \right] \nonumber \\
\PL(t_i) &=& \mathbb{E}_i \left[ \sum_{T_n > t_i} \Delta_n
B(t_i,T_{n})  N_{[K_d,K_u]}(T_{n})  \right] \eea
where the tranche loss, $L_{[K_d,K_u]}$, and tranche notional,
$N_{[K_d,K_u]}$, have been defined in section~\ref{sec:pricing}.
Here $ \Delta_n $ is the day count fraction for the period
$[T_{n-1},T_n]$ and $ \mathbb{E}_i $ stands for the expectation
value conditional on the history up to time $ t_i $. Assuming that
the last time grid point falls on a coupon payment date, we impose
the following boundary conditions for $ \DL $ and $ \PL $  (here $ i
$ is the $N$-index and $ n $ in a $Y$-index):
\bea \label{boundaryConditions}
\DL_{i,n}(t_M) &=& 0 \nonumber \\
\PL_{i,n}(t_M) &=& \Delta_{M}  N_{[K_d,K_u]}(t_M)  \eea

Using the tower law for conditional expectations \beq \label{tower}
\mathbb{E}_t \left[ X_T \right] \equiv \mathbb{E} \left[ \left. X_T
\right| \mathcal{F}_t \right] = \mathbb{E} \left[ \left. \mathbb{E}
\left[ \left. X_T \right| \mathcal{F}_s \right] \right|
\mathcal{F}_t \right] \; \; , \; \; t \leq s \leq T \eeq and
splitting off the first terms  of the sums in (\ref{DLPL}), we
obtain the recursive formulas
\bea \label{recursiveDLPL} \DL(t_i) &=& B(t_i,t_{i+1} ) \mathbb{E}_i
\left[ \DL(t_{i+1}) \right] + B(t_i,t_{i+1}) \left(
\mathbb{E}_i \left[ L_{[K_d,K_u]}(t_{i+1}) \right] - L_{[K_d,K_u]}(t_i) \right) \nonumber \\
\PL(t_i) &=& B(t_i,t_{i+1} ) \mathbb{E}_i \left[ \PL(t_{i+1})
\right] + \Delta_{n } N_{[K_d,K_u]}(T_{n})  \delta_{t_i, T_{n}} \eea
Here the second term in the second equation is non-zero only on
coupon payment dates, and corresponds to a coupon added on that
date. Using the one-step backward equations to evaluate the
expectations entering (\ref{recursiveDLPL}), we obtain the backward
induction pricing method for a tranche.

\subsection{Tranche option}

Given the algorithm for calculating the tranche mark-to-market by
backward recursion on a tree, we can price a tranche option using
the standard tree option pricing procedure. Note that today's value,
$V_0$ of the tranche option with strike $k$ and exercise date $T_1$
and maturity $T_2$ is given by:
\beq V_0 =\mathbb{E}_0[\MTM_{T_1}(T_2,k)^+] \eeq
where $\MTM_t(T,c)$ is the mark-to-market at time $t$ of the
underlying tranche paying coupon $c$ and with maturity $T$.
Therefore, to price the tranche option, we build a tree or lattice
up to maturity $ T_2 $, and then roll the tranche price backward on
this tree from $ T_2 $ to $ T_1 $. This provides the boundary
condition at $ T_1 $ for the tranche option which is then rolled
backward in time from $ t = T_1 $ to $ t = 0 $.

\subsection{Forward starting tranche}

Forward starting tranches provide protection against tranche losses
in a pre-specified future period $ [t,T] $. The distinguishing
feature is that defaults occurring prior to $t$ do not affect the
subordination of the tranche\footnote{Hence the forward tranche is
not simply the difference of two standard tranches with maturities
given by $t$ and $T$ (the latter is sometimes referred to as the plain vanilla
forward tranche).}. In particular, a forward tranche with low
strike $K_d$ and high strike $K_u$ can be valued as the forward
value of a tranche with \emph{adjusted} strikes, $K'_d$ and $K'_u$:
\beq \label{forwardTranche} V_0 = \mathbb{E}_0[\MTM_t(T; K'_d,
K'_u)] \eeq
where $\MTM_t(T; K_u ,K_d)$ is the mark-to-market at time $t$ of a
tranche with maturity $T$ and low and high strikes $K_d$ and $K_u$.
The strikes are adjusted by the loss, $L_t$, at time $t$  and given
by: $K'_u = \min(1, K_u+L_t)$ and $K'_d = \min(1, K_d+L_t)$. This
dependence of the payoff on the loss makes the forward tranche path
dependent.

To price this on a tree we need to know the mark-to-market,
$\MTM_i(t)$, of the tranche at all loss nodes, $i$, of the tree at
time $t$. The payoff of the forward tranche is dependent on the loss
at $t$ and hence each $\MTM_i(t)$ needs to be calculated on a
separate sub-tree emanating from node $i$. The values $\MTM_i(t)$
provide a boundary condition for the tree at time $t$ which can be
rolled back to $t=0$ as usual\footnote{This pricing
algorithm can be somewhat simplified if the forward induction-based
calibration of section~\ref{BJN} is used. In this case, marginal
default probabilities are calculated at the calibration stage.
Hence, we only need to roll the tranche MTM backward in time from $
T $ to $ t $. The price at time $ t = 0 $ is then given by a
weighted sum of $ \MTM_i(t) $ at all nodes $ i $, with the weights
being the state probabilities at time $ t $. Note that the same
comment applies to tranche options as well.}.

\subsection{Leveraged super senior}

Leveraged Super Senior (LSS) is a tranche with the added feature
that the trade knocks out once a certain trigger is breached.
Different versions  define the trigger to be either the portfolio
loss or spread, or MTM of the tranche. In particular, for the loss
trigger, the trade knocks out when the portfolio loss $L_t$ hits a
(deterministic) barrier $ \bar{L}(t) $ for the first time. Assuming
a deterministic and fixed recovery $ R $ as before, we can translate
this into an equivalent default count boundary $ \bar{N}(t) =
\bar{L}(t)/(1-R) $. The random hitting time $ \tau $ is therefore
the following:
\beq
\label{tau} \tau = \inf \{ t: N_t \geq
\bar{N}(t) \}
\eeq
The payoff for the protection buyer is given by:
\beq
\label{LSSpayoff}
P(\tau) = \min \left( K, \MTM(\tau + \Delta t) \right) =
\MTM(\tau + \Delta t) - \left( \MTM(\tau + \Delta t) - K \right)^{+}
\eeq
where $ K $ is the collateral posted and $ \Delta t $ is the unwinding period (typically
two weeks) during which the trade is terminated and unwound after first breaching the barrier.
%
 %
%
%
Equation~(\ref{LSSpayoff}) implies that the LSS can be viewed as a
portfolio of a long position in the super-senior tranche and a short
position in an American barrier option (the ``gap risk option'').
Let $ C(t) $ be the price of this option at time $t$. Assuming for
simplicity that the length of the unwinding period $ \Delta t $ is
equal to the time step on the tree, the option can be priced using
the standard backward recursion:
%
%
\bea \label{gapRisk} C(t_i) &=& B(t_i,t_{i+1} ) \mathbb{I}_{ N_{t_i}
< \bar{N}(t_i)}
\mathbb{E}_i \left[ C(t_{i+1}) \right] \nonumber \\
&+& B(t_i,t_{i+1}) \mathbb{I}_{N_{t_i} \geq \bar{N}(t_i)}
\mathbb{E}_i \left[ \left( \MTM(t_{i+1}) - K \right)^{+} \right]
\eea
%

\section{Numerical results}
\label{NumResults}

In this section, we present results obtained with BSLP based on
calibration to \emph{iTraxx Europe Series 6} tranche quotes from
March 15th, 2007. We look at two specifications of the model:
\begin{itemize}
\item[{\bf A}:] The $0\%$ $Y_t$-volatility case with
mean reversion speed $ a = \infty $ and  transient contagion $\gamma = 0$. This
is equivalent to the local intensity model
presented in Sect.~\ref{DTM}.
\item[{\bf B}:] A log-normal specification for the $Y_t$ process (\ref{logNormal}) with
volatility $\sigma = 70\%$, mean reversion speed $a = 30\%$  and
transient contagion $\gamma = 0$.
\end{itemize}

\subsection{Calibration}

\begin{figure}[t]
\begin{center}
\includegraphics{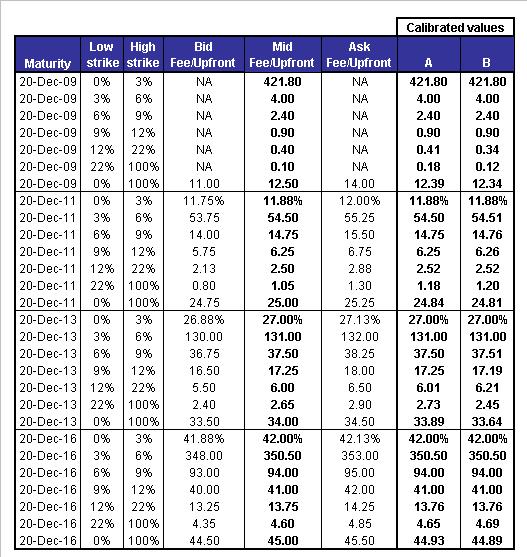}
\caption{Calibration results for BSLP. Note that quotes are given as
spreads in bps except for the 5,7, and 10y equity tranches which are
quoted as upfront in percent (with a coupon of 500bps). Calibration
is done to mid quotes. Indicative bid/ask quotes are included for
reference to assess calibration quality. Note that 3y bid/ask
spreads are not available. } \label{fig:calibration}
\end{center}
\end{figure}

Calibration of BSLP to the tranche quotes was done using the 2D
calibration described in Sect.~\ref{calibration}. We present the
quality of the calibration in Fig.~\ref{fig:calibration} where we
indicate mid as well as bid and ask levels. We see that in all cases
the calibrated values lie well within the bid/ask spreads.

\subsection{Pricing}
\subsubsection{Tranchlets}
\begin{figure}[t]
\begin{center}
\includegraphics[width=16cm]{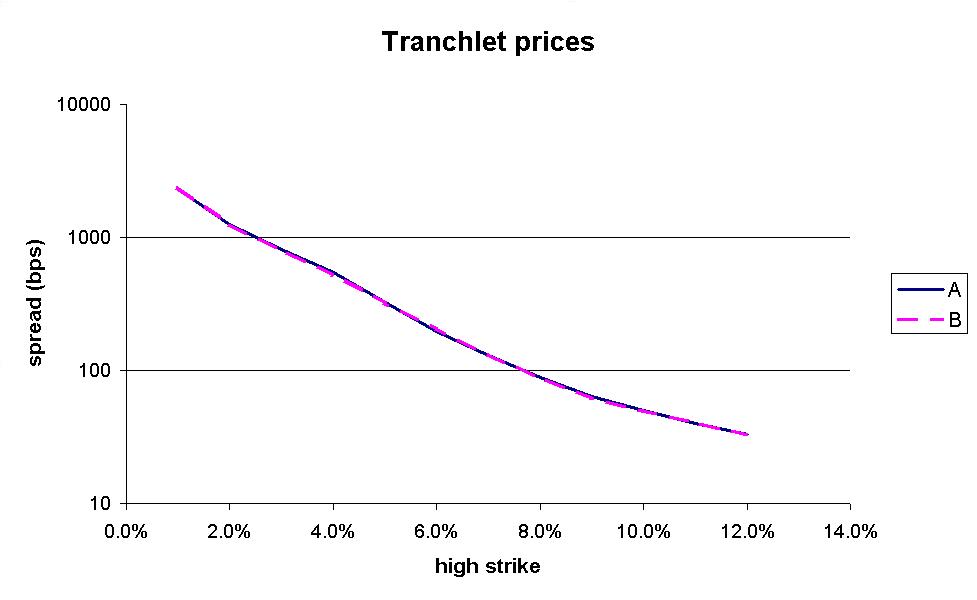}
\caption{Tranchlet ($1\%$-wide) prices for models A and B on
logarithmic scale. Maturity is the 20th December 2016.}
\label{fig:tranchlets}
\end{center}
\end{figure}

Next we investigate the behavior of the two calibrated BSLP models A
and B. First we look at implied tranchlet pricing, i.e. prices of
thin CDO tranches on the iTraxx portfolio. In particular we look at
the 10 year $1\%$-wide tranches up to $12\%$ and plot the results in
Fig.~\ref{fig:tranchlets}.

The first observation is that the tranchlet curve is close to linear
on a logarithmic scale, resulting in a smooth interpolation and
extrapolation to the equity region.

The second observation is that the $0\%$ and $75\%$  volatility
cases produce nearly identical curves. This is a reassuring result
that suggests that the local intensity model is indeed sufficient
for tranche price interpolation. Note that this is not a completely
trivial observation. Indeed, similarly to regular tranches,
tranchlet prices are determined by a set of marginal loss
distributions at a set of horizons. However, market tranche quotes
do not uniquely define those marginal probabilities; they only
provide integral constraints on them (i.e.\ the market is incomplete
- see also the remarks at the end of section~\ref{BJN}). When we
calibrate BSLP with volatilities $0\%$ and $75\%$, strictly speaking
we end up with two different sets of marginal loss distributions,
whose difference could in principle amplify when projected onto thin
tranchlets. Thus our findings imply that volatility assumptions have
a negligible impact on marginal loss distributions even for
incomplete markets.



\subsubsection{Tranche Options}
The main aim of the BSLP model is to price exotic credit
derivatives. Here we look at the tranche option as an example. We
focus on  5y into 5y options, i.e.\ exercise date 20-Dec-2011 and
maturity 20-Dec-2016. The tranches we consider here are the index
($0\% - 100\%$) and the equity ($0\%-3\%$) tranche.

\begin{figure}[t]
\begin{center}
\includegraphics[width=16cm]{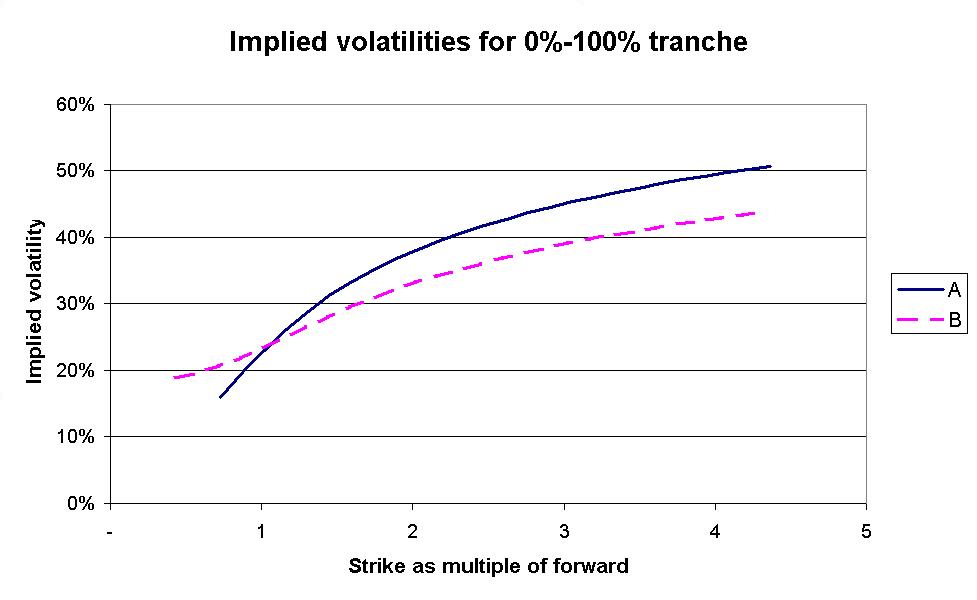}
\caption{Implied volatilities for index ($0\%-100\%$ tranche)
option. Exercise date is 20-Dec-2011 and maturity is 20-Dec-2016.
Strikes are given as a multiple of the forward value, which is at
69bps. } \label{fig:optIndex}
\end{center}
\end{figure}

\begin{figure}[h]
\begin{center}
\includegraphics[width=16cm]{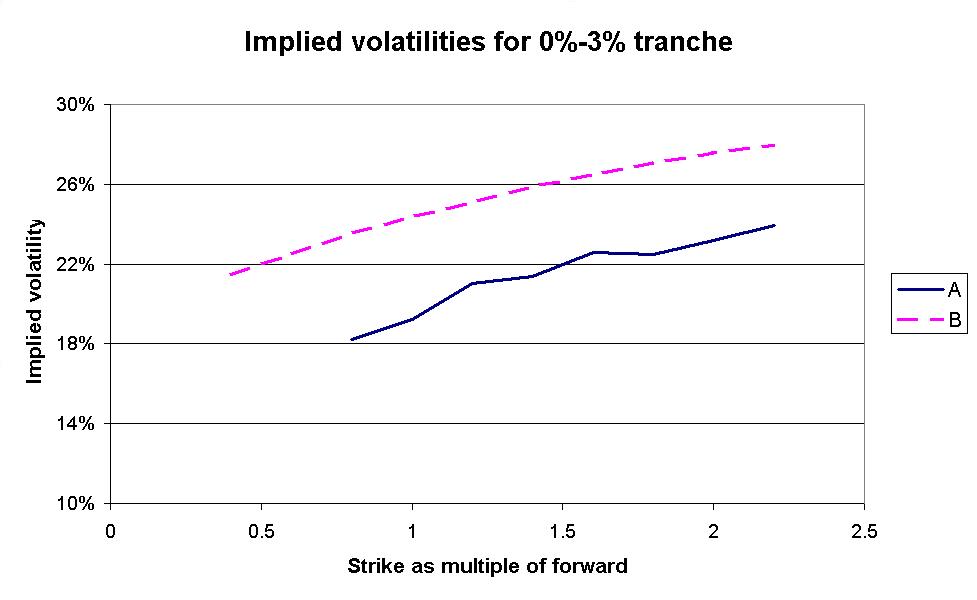}
\caption{Implied volatilities for equity ($0\%-3\%$ tranche) option.
Exercise date is 20-Dec-2011 and maturity is 20-Dec-2016. Strikes
are given as a multiple of the forward value, which is at 2725bps. }
\label{fig:optEquity}
\end{center}
\end{figure}

Figs.~\ref{fig:optIndex} and~\ref{fig:optEquity} show the
implied (Black) volatilities for these options for both model
choices A and B.

We start by discussing the index option (Fig.~(\ref{fig:optIndex})).
It can be seen clearly that for both models A and B there is a
strong skew in the implied volatilities.

 It should be noted that even for model A which has
$0\%$ volatility for the portfolio intensity the implied
volatilities are high (the ATM vol is around $22\%$). As we will see
in the next section, this is because
the model gives rise to a  non-zero dispersion of forward spreads even in the absence of
Y-volatility.

One can also see that adding non-zero $Y$-volatility results in a
rotation of the volatility smile curve, i.e. volatilities increase
for low strikes and decrease for high strikes.  Again, this can be
understood by looking at the behavior of the conditional forward
spreads as discussed in section~\ref{condForwards}.

For the equity option (Fig.~(\ref{fig:optEquity})) we note that
increasing $Y$-volatility results in an upward  parallel shift of
the smile curve. We can also see that the curve for the local
intensity model (A) is not very smooth. This is a result of the fact
that the local intensity model is fundamentally discrete, i.e.\ the
dynamic default variable can only take on integer values. This means
in turn that the distribution of conditional forwards is also
discrete. As a result option prices are linear for strikes that lie
between the conditional forward levels.

\subsubsection{Conditional Forwards}
\label{condForwards}

A key quantity for assessing dynamics of a multi-period model is the
forward spread distribution. Here we focus on the 5 year spreads,
conditional on the default level at 20-Dec-2011.

These spreads tell us how the model links spread and default
dynamics. This can be  compared to intuition on how risk will be
priced in different (default) states of the world. Of course these
forwards will also play a direct role in exotic pricing, e.g.\ a
tranche option can be viewed as essentially an option on the forward
spread.

\begin{figure}[t]
\begin{center}
\includegraphics{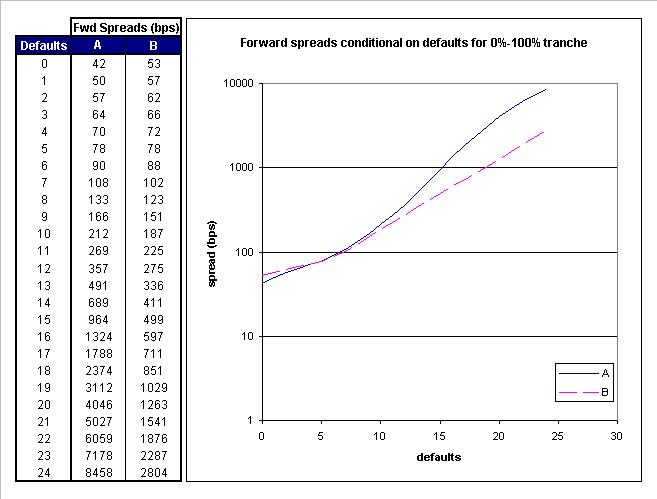}
\caption{Forward spreads on a logarithmic scale conditional on the
default level for the index tranche for models A and B. Start date
is 20-Dec-2011 and maturity is 20-Dec-2016. The unconditional
forward level is ca. 69bps. }
 \label{fig:condFwdIndex}
\end{center}
\end{figure}

\begin{figure}[h]
\begin{center}
\includegraphics{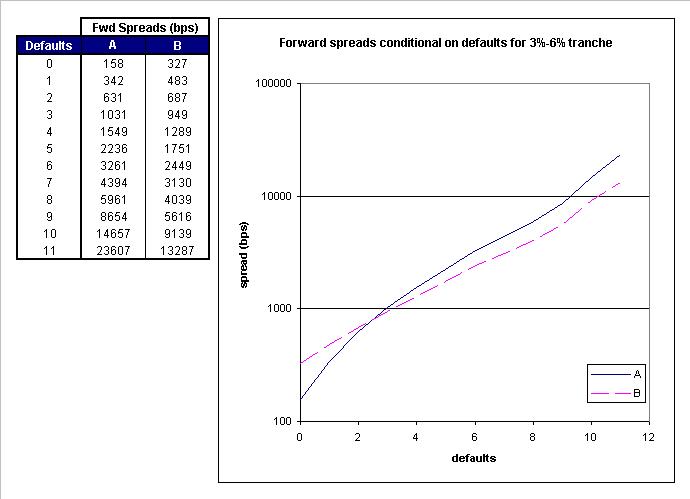}
\caption{Forward spreads on a logarithmic scale conditional on the
default level for $3\%-6\%$ tranche for models A and B. Start date
is 20-Dec-2011 and maturity is 20-Dec-2016. The unconditional
forward level is  750bps. Note that the tranche is wiped out after
11 defaults.} \label{fig:condFwdMezz}
\end{center}
\end{figure}

The forwards levels are plotted in Figs.~\ref{fig:condFwdIndex}
and~\ref{fig:condFwdMezz} for \ $0\%-100\%$ and a $3\%-6\%$ tranche
respectively.

First we should note that $0\%$ volatility (model A) does not imply
0 variance for the distribution of conditional forwards. This leads
to non-zero implied tranche option volatilities as described in the
previous section. The reason for this is that even though there is
no spread diffusion, we still have spread jumps  determined by the
contagion factors.

One can also see a rapid increase in spread levels as defaults
increase, which indicates that the model implies a high level of
contagion. An interesting feature is that the increase in forward
spreads is less pronounced for model B than A, i.e.\ introducing
$Y$-volatility reduces the level of contagion.

This can be understood since we are replacing spread contagion with
spread diffusion while calibrating to the same portfolio loss
distribution. Another way to see this is to note that adding spread
diffusion will decorrelate losses and spreads.

This is an important  effect, which is likely to have significant
impact on exotic pricing.

\subsection{Risk}
In this section we explore the first order (spread) risk in the BSLP
model. Since BSLP is driven by a stochastic $Y$-process, it is
natural to look at deltas with respect to
moves in todays value $ Y_0 $. The meaning of this procedure can be
clarified using Eq.(\ref{lambdaS}): if we keep contagion factors $
F(N_k,0)$ constant and shift $ Y_0 $, this is equivalent to shifting
the index swap intensity $ \lambda_s(0) $, and thus can be used to
specify the index delta.

We note that, the same Eq.(\ref{lambdaS}) also shows that shifting $
Y_0 $ while keeping contagion factors $ F(N_k,0)$ constant is
equivalent to a common rescaling of all contagion factors.

\subsubsection{Tranche Deltas}
First we look at how tranche prices behave as we shift the initial
$Y$-value, $Y_0$. In particular, we can look at the hedge ratio to
the index.

Let us denote the mark-to-market of a tranche with low strike $k$,
high strike $l$ and coupon $c$ as a function of $Y_0$ by:
$\MTM_{[k,l]}(c, Y_0)$. The par spread of this tranche is
$S_{[k,l]}$, defined by: \beq \MTM_{[k,l]}(S_{[k,l]}, Y_0) = 0 \eeq

For a given tranche with strikes $k$ and $l$, we now define the
tranche delta $\Delta_{[k,l]}$ as follows:
\beq \Delta_{[k,l]} \equiv \frac{\MTM_{[k,l]}(S_{[k,l]},Y_0+
\epsilon) - \MTM_{[k,l]}(S_{[k,l]},Y_0)}{\MTM_{[0,1]}(S_{[0,1]},Y_0+
\epsilon) - \MTM_{[0,1]}(S_{[0,1]},Y_0)} =
\frac{\MTM_{[k,l]}(S_{[k,l]},Y_0+
\epsilon)}{\MTM_{[0,1]}(S_{[0,1]},Y_0+ \epsilon) } \eeq

\begin{figure}[t]
\begin{center}
\includegraphics{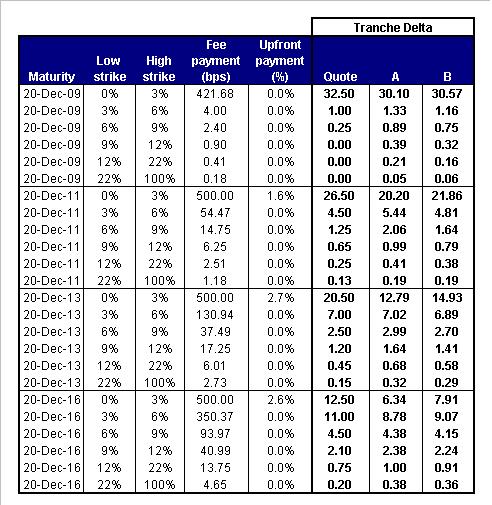}
\caption{Tranche deltas computed with models A and B compared to
iTraxx quoted deltas.} \label{fig:trancheDelta}
\end{center}
\end{figure}

For index tranches these deltas can be compared to the quoted spread
deltas. This is done in Fig.~\ref{fig:trancheDelta}. We see that
deltas are reasonably close to the quoted values. BSLP deltas tend to
be lower in the equity region and higher for super senior tranches.
We note also that adding $Y$-volatility (model B) tends to improve
the delta match.

\subsubsection{Option Deltas}
We now look at deltas of tranche options. Again we consider the
impact of a shift in initial $Y$-level and compute the option delta
as the change in mark-to-market of the option divided by the change
in mark-to-market of the forward tranche. In other words we are
computing the hedge ratio of the tranche option with respect to a
position in the underlying forward.

We can compare this to the hedge ratio obtained by using the Black
formula. In this case we first calculate the implied Black
volatility of the option. After shifting $Y_0$ we get new forward
levels and risky annuities for the underlying tranche. Using these
shifted forwards and annuities but keeping the implied volatility
constant we can compute a new option price using the Black formula.
This gives rise to a Black delta or hedge ratio.

\begin{figure}[t]
\begin{center}
\includegraphics[width=16cm]{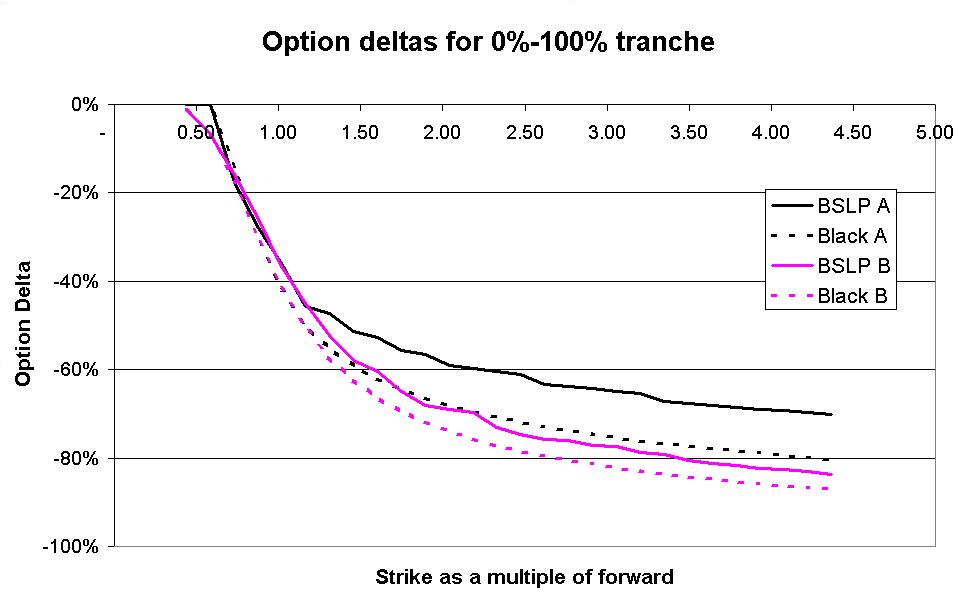}
\caption{Option deltas for the index tranche computed with models A
and B compared to Black option deltas. Strikes are given as
multiples of the forward level which is ca. 69bps.}
\label{fig:optDeltaIndex}
\end{center}
\end{figure}

\begin{figure}[h]
\begin{center}
\includegraphics[width=16cm]{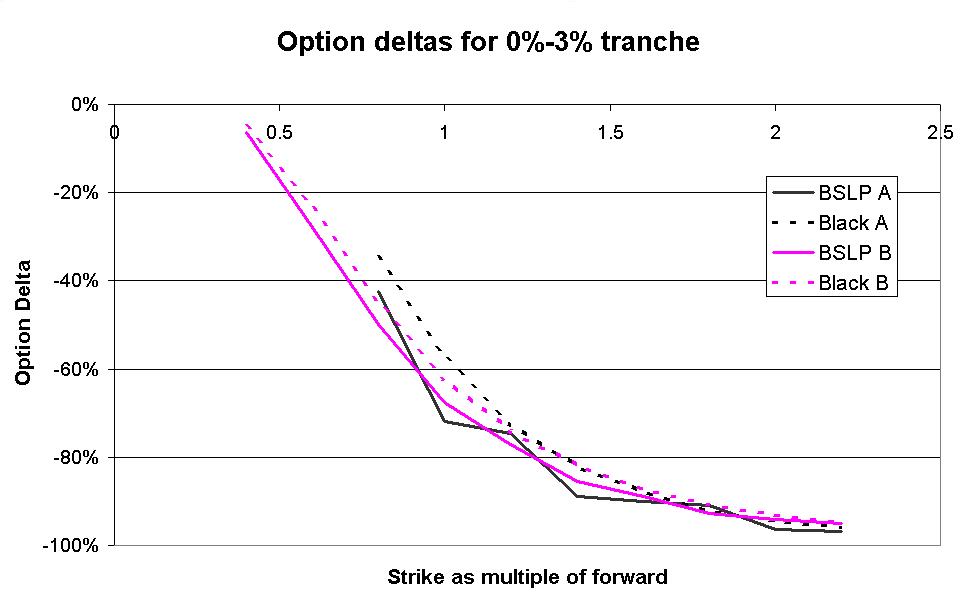}
\caption{Option deltas for the $0\%-3\%$ equity tranche computed
with models A and B compared to Black option deltas. Strikes are
given as multiples of the forward level which is ca. 2725bps.}
\label{fig:optDeltaEquity}
\end{center}
\end{figure}

The results are given for a $0\%-3\%$ equity tranche option and the
$0\%-100\%$ index option in Figs.~\ref{fig:optDeltaEquity}
and~\ref{fig:optDeltaIndex} respectively. As before the exercise
date of the options is 20-Dec-2011 and the maturity is 20-Dec-2016.

We can see that BSLP and Black deltas are of the same order. The
match is better for the equity option than the index option,
especially for high strikes. We note also that Black and BSLP deltas
are closer to each other for model specification B, i.e. with
non-zero $Y$-volatility.

As in the case of the implied Black volatility smile we can see that
the curve of option deltas as a function of strike is not very
smooth for the $0\%$ volatility case. Again this is due to the
fundamental discreteness of the local intensity model.

\section{Summary}
\label{Conclusion}

BSLP is a  model of the portfolio loss and loss intensity. It can be
used for the pricing and risk management of vanilla as well exotic
credit derivatives which depend on the portfolio loss.

Because the model is low-dimensional and Markovian, efficient
lattice or tree implementations are possible. In addition to such
numerical methods, analytical approaches such as the adiabatic
approximation can be used. A possible extension to the model
considers multi-factor generalizations of the driving stochastic
process.

BSLP explicitly models credit contagion and provides insight on how
this impacts the forward loss dynamics. This is a key for the pricing
and risk management of exotic portfolio credit derivatives.

The model achieves a near perfect calibration to any set of
portfolio tranche quotes due to a semi-parametric representation of
the contagion factors. These quotes can either be obtained from the
market for liquid indices or from an underlying bespoke tranche
pricing model.

Efficient lattice or tree pricing algorithms for the model have been
developed, including  fast calibration algorithms based on a 1D
projection of the full 2D model. Tree pricing is most suitable for
pricing  exotic credit portfolio derivatives that incorporate
optionality and/or weak loss-path dependence such as tranche
options, forward tranches, leveraged super-senior etc.. However,
Monte Carlo implementations of the model (not discussed in this
paper) are also possible in order to price more strongly path
dependent products.

The calibrated model retains flexibility in adjusting the dynamics
by choosing a concrete specification of the underlying stochastic
driving process. This provides some freedom in the pricing of exotic
derivatives. The local intensity limit of BSLP is a model in its own
right that can be used to price non-standard tranches by
arbitrage-free interpolation.

\clearpage
\appendix
\def\theequation{\thesection.\arabic{equation}}

\section*{Appendix A: Swap intensity and NtD intensity}

\def\thesection{A}
\setcounter{equation}{0}

Here we formally derive the relation (\ref{lambdaS}) between the NtD
intensities specifying the transition probabilities  (\ref{ntD}) in
the full 2D model, and the stochastic index swap intensity $
\lambda_s (t) $.

The swap intensity is defined as the intensity that prices the index
correctly as a single name CDS. More precisely, we want the
following. Given a stochastic  intensity $\lambda_s(t)$ we denote
the distribution, $Q(t,T)$, of the first jump $\tau$ of the
associated counting process by:
\beq Q(t,T) \equiv P[\tau > T |\mathcal{F}_t]  \eeq
This is all we need to price a CDS. To ensure that the value of this
CDS corresponds to the value of the index calculated using the
portfolio default process $N_t$ we need:
 \beq \label{lambdaS2} Q(t,T) = \frac{ \mathbb{E}
\left[ N-N_T | \mathcal{F}_t \right]}{ N-N_t } \eeq
The intensity  corresponding to $Q(t,T)$ is given by:
\beq \lambda_s(t) = - \left. \frac{\partial Q}{\partial T} \right
|_t (t,T)\eeq
Using equation~(\ref{lambdaS2}) this gives:
\beq \label{firstorder} \lambda_s
(t) dt =
\frac{\mathbb{E} [ N_{t+dt} | \mathcal{F}_t ] - N_t}{ N - N_t}
\eeq
We can evaluate the conditional expectation in (\ref{firstorder}) as
follows: \beq \label{condExp} \mathbb{E} [ N_{t+dt} | \mathcal{F}_t
] = \sum_{k = 0}^{ N - N_t} ( N_t  + k) P[N_{t+dt} = N_t +k|
N_t,Y_t,t] \eeq where $ P[N_{t+dt} = N_t +k| N_t,Y_t,t] $ is the
probability of the transition $ N_t \rightarrow N_{t+ dt} = N_t + k
$. As we only allow for at most one step transition in the
infinitesimal time $ dt $, the sum in (\ref{condExp}) reduces to
just two terms: \beq \label{twoterms} \mathbb{E} [ N_{t+dt} |
\mathcal{F}_t ] = (N_t + 1) \lambda_{NtD}(N_t,Y_t,t) dt + N_t \left(
1 - \lambda_{NtD}(N_t,Y_t,t) dt \right) \eeq Substituting this
relation into (\ref{firstorder}), we arrive at the sought-after
relation between the two intensities: \beq \label{sought}
\lambda_s(N_t,Y_t,t) = \frac{ \lambda_{NtD}(N_t,Y_t,t)}{ N - N_t}
\eeq

Note that $\lambda_s(t)$ is also the average of the portfolio single
name intensities $\lambda_i$. This follows since in the absence of
simultaneous defaults we know that the portfolio default intensity
$\lambda_{NtD}$ is just given by the sum of single name intensities.
In other words:
\beq \lambda_{NtD} =  \sum_{i=1}^{N-N_t} \lambda_i =
(N-N_t)\lambda_s \eeq

\appendix
\def\theequation{\thesection.\arabic{equation}}

\def\thesection{B}
\setcounter{equation}{0}

\section*{Appendix B: BSLP in continuous time}

Here we present a continuous-time formulation of the BSLP model for
a jump-diffusion specification of the driving $ Y $-process given in
Eq.(\ref{SDEY}). (A continuous-time formulation with a discretized $
Y$-variable is further analysed in Appendix C).
In this appendix, we will use a more conventional notation $ (x,y) $
instead of $ (N,Y) $ for the dynamic variables of the BSLP model,
with $ x $ denoting the defaulted fraction $ N_t /N $ instead of the
absolute default level $ N_t $.


The general form of the infinitesimal jump-diffusion generator $
\mathcal{L} $ acting on the pdf $ P(\vec{z}',t| \vec{z},s) $ for
$n$-dimensional  random variables $ \vec{z}', \, \vec{z} \in R^n $
is as follows \bea \label{jumpdiffGen} \mathcal{L} P(\vec{z}',t|
\vec{z},s) & = & \sum_{i = 1}^{n} \mu_i(\vec{z},s) \frac{\partial P
(\vec{z}',t| \vec{z},s) }{\partial z_i} + \sum_{i,j = 1}^{n}
\frac{1}{2}  \sigma_{ij}^2(\vec{z},s)
\frac{\partial^2 P (\vec{z}',t| \vec{z},s)}{\partial z_i \partial z_j}  \nonumber \\
& + &  \sum_{\vec{z}'' \neq \vec{z}} W(\vec{z}''| \vec{z},s) \left[
P(\vec{z}',t| \vec{z}'',s) - P(\vec{z}',t| \vec{z},s) \right]
\eea
while the adjoint operator $ \mathcal{L}^{*} $ reads\footnote{Recall the
definition of adjoint operator $ \mathcal{L}^{*} $:
\[
\int dx \, P_1 \mathcal{L}[P_2] = \int dx \, P_2 \mathcal{L}^{*} [P_1]
\]
where $ P_1(x), \, P_2(x)$ are two arbitrary probability densities.}
\bea
\label{jumpdiffGenStar}
\mathcal{L}^{*} P(\vec{z}',t| \vec{z},s) & = & - \sum_{i = 1}^{n}
\frac{\partial}{\partial z_i'}  \left[ \mu_i(\vec{z}',t) P(\vec{z}',t| \vec{z},s) \right]
+ \sum_{i,j = 1}^{n} \frac{1}{2}  \frac{\partial^2}{ \partial z_i' \partial z_i' }
\left[\sigma_{ij}^2(\vec{z}',t) P (\vec{z}',t| \vec{z},s) \right] \nonumber  \\
&+&  \sum_{\vec{z}'' \neq \vec{z}}  \left[W(\vec{z}' | \vec{z}'',t) P(\vec{z}'',t| \vec{z},s)
 -  W(\vec{z}'' | \vec{z}',t)
P(\vec{z}',t| \vec{z},s) \right] \eea Here $ W(\vec{z}' | \vec{z},t)
$ is the jump measure determining the jump probability  in time $ dt
$ together with the jump size distribution.  Using standard
conventions, we write \beq \label{W} W(\vec{z}' | \vec{z},t) =
\lambda(\vec{z},t) w(\vec{z}'- \vec{z} | \vec{z}, t) \eeq where $
\lambda(\vec{z},t) $ stands for the jump rate (intensity) and $ w(
\delta \vec{z} | \vec{z},t) $ is a pdf of the jump size $ \delta
\vec{z} $ given the initial point $ \vec{z} $. The forward and
backward equations are \beq \label{forwardBackward} \frac{ \partial
P(\vec{z}',t| \vec{z},s)}{ \partial t} = \mathcal{L}^{*}
P(\vec{z}',t| \vec{z},s)  \; \; , \; \; \frac{
\partial P(\vec{z}',t| \vec{z},s)}{\partial s} = - \mathcal{L}
P(\vec{z}',t| \vec{z},s) \eeq We note that the generator can be
written in the following form \beq \label{formOfGenerator1} A =
\mathcal{L}_0 + \mathcal{L}_1 \eeq where $ \mathcal{L}_0 $ and $
\mathcal{L}_1 $ correspond to diffusion and jump parts of the
generator, given by the first two terms and the last term in
(\ref{jumpdiffGen}), respectively.


Let us now define the operators $ \mathcal{L}_0 $ and $
\mathcal{L}_1 $ in our specific 2D setting with $ \vec{z} = (x,y) $.
In the BSLP model, the generator $ \mathcal{L}_0 $ acts only on $ y
$. We therefore set \beq \label{L0} \mathcal{L}_0 P(x,y,t|x_0,y_0,s)
 =   \mu(y_0,s)
\frac{\partial   }{\partial y_0} P(x,y,t| x_0,y_0,s)
+ \frac{1}{2}  \sigma^2(y_0,s)
\frac{\partial^2 }{\partial y_0^2} P (x,y,t| x_0,y_0,s)
\eeq where the functions $ \mu(y,t) $ and $ \sigma (y,t) $ are
defined in the SDE (\ref{SDEY}). For the adjoint operators we obtain
\beq \label{adjoint} \mathcal{L}_0^{*} P(x,y,t|x_0,y_0,s)
 =   -
\frac{\partial}{\partial y}  \left[ \mu(y,t) P(x,y,t| x_0,y_0,s) \right]
+ \frac{1}{2}
\frac{\partial^2 }{\partial y^2} \left[\sigma^2(y,t)
 P (x,y,t| x_0,y_0,s) \right]
\eeq The second generator $ \mathcal{L}_1 $ corresponds to the case
where jumps $ x \rightarrow x + \Delta x $ are enabled, and in
addition are accompanied by a related jump in $ y $ (which arises
due to the $ d N_t$-term in (\ref{SDEY})). The rate of this process
is proportional to the product of the $ y $-variable and the
fraction of surviving names, times the contagion factor $ F(x) =
q(x) f(x) $, similarly to the treatment above: \beq
\label{intensityJointJump} \lambda (x,y,t) =  y (1 - x) q(x) f(x)
\eeq and the jump size distribution is a product of two
delta-functions: \beq \label{jumpmeasure} w_1(x- x_0,y- y_0|
x_0,y_0,s) =  \delta(x - x_0 - \Delta x) \delta(y - y_0 - \Delta y)
\eeq Using (\ref{jumpmeasure}) and (\ref{intensityJointJump}), we
calculate the generator $ \mathcal{L}_1 $ \bea \label{L1jump}
\mathcal{L}_1 P(x,y,t| x_0,y_0,s) & = &  \lambda(x_0,y_0,s)
 \left[ P(x,y,t| x_0 + \Delta x, y_0 + \Delta y,s) \right. \nonumber \\
& - &  \left. P( x, y,t| x_0,y_0,s) \right]
\eea
and for the adjoint operator we have
\bea
\label{Lstar1} \mathcal{L}_1^{*} P(x,y,t| x_0,y_0,s)
& = &  \lambda(x - \Delta x, y - \Delta y,t)
P(x - \Delta x,y - \Delta y ,t| x_0, y_0,s)   \nonumber \\
& - & \lambda(x,y,t)  P(x,y,t| x_0,y_0,s) \eea To summarize, $
\mathcal{L}_0 $ in (\ref{formOfGenerator1}) collects terms in the
generator $ \mathcal{L} $ where $ Y_t $ serves as the only dynamic
variable while the default counting variable $ N_t $ enters as a
parameter. The second term $ \mathcal{L}_1 $ describes the part of
the generator that contains both $ Y_t $ and $ N_t $  as dynamic
variables. This special structure of the generator can be used to
set up an adiabatic perturbation theory for the model (see Appendix
D).

\appendix
\def\theequation{\thesection.\arabic{equation}}

\def\thesection{C}
\setcounter{equation}{0}

\section*{Appendix C: Discretized BSLP in continuous time}

Here we  discuss how the continuous-time formalism presented in
Appendix B
 changes if we keep time continuous but
assume that the $Y_t $-variable can only take discrete values from
some finite set. Such analysis may be of interest as in this case
the model becomes that of a 2D Markov chain, allowing powerful
methods of  Markov chain theory to be used to build fast and
accurate numerical approximations to the dynamics of the original
model, get more insight into the model behavior, and build
connections to previous research. In particular, we will use this
reformulation to compare our approach with the Davis-Lo model
\cite{DL}, as well as establish a link with a general and rich class
of quasi-birth-death (QBD) processes.

Unlike the setting of Appendix B, which uses a jump-diffusion
framework, here we assume that the $ Y_t $-variable is discretized.
 For the general case of a two-dimensional
process in the $ NY$-plane, the states are parameterized by two
indices $ (i,n) $ ($ i $ for number of defaults, and $ n $ for
$Y$-state). Correspondingly, the 2D generator carries four indices
instead of two. We'll use the notation $ A_{in|jm}(t) $ for the
matrix elements of the generator. In matrix notation, the generator
can be viewed as a block matrix \bea \label{2D} A \, = \, \left(
\begin{array}{clcrcc}
L^{(0)}  & F^{(0)} & 0 & 0 & \cdots & 0  \\
0 & L^{(1)} & F^{(1)}  &  0 & \cdots & 0 \\
0 &  0      & L^{(2)} & F^{(2)}  & \cdots & 0 \\
\vdots                        \\
 0  & 0 & 0 &  0 & \cdots & 0
\end{array} \right)
\eea where all matrices $ L^{(i)},\, F^{(i)} $ have dimension $ M
\times M $, with $ M$  being the dimension of the $ Y$-space. The
interpretation of these matrices is as follows. The matrix $ L^{(i)}
$ gives intensities of transitions between $ Y$-states when there is
no change of the $ L $-state during the infinitesimal time step $
\Delta t $, while the matrix $  F^{(i)} $ provides intensities of
joint events of a jump in the loss variable during the interval $
[t, t + dt] $ accompanied by a transition between $ Y $-states. As
usual, all off-diagonal elements should be positive, diagonal
elements should be negative, and each row in $ A $ should sum up to
zero\footnote{Such a block-matrix structure of the generator is
characteristic of the so-called Quasi-Birth-Death (QBD) processes.
In the terminology of QBD, the loss variable is the ``level'' while
the Y-variable is the ``phase''. Symbols ``L'' and ``F'' stand for
``local'' (without change of level) and ``forward'' (level is
changed by one unit), respectively.}.

Given the 2D generator matrix $ A$, the forward equation takes a block-matrix form
\beq
\label{forward2D}
\frac{ \partial P(t,T)}{ \partial T}  =   P(t,T) A
\eeq
where $ P(t,T) $ has matrix elements $ P_{in|jm} (t,T) $.

The Davis-Lo model \cite{DL} is a particular case of a QBD credit
process where the dynamics of $ Y$ is locally deterministic. Matrix
elements of $ A $ are given in this model by $ 2 \times 2 $ matrices
\bea \label{DavisLo} L^{(i)} = \, \left( \begin{array}{cc}
- \lambda ( N - i) & 0 \\
\mu & - \mu - a \lambda ( N - i)
\end{array} \right)
\; \;
F^{(i)} = \, \left( \begin{array}{cc}
0  & \lambda ( N - i)  \\
0 &  a \lambda ( N - i)
\end{array} \right)
\eea Note that the form of  $ F^{(i)} $ implies that whenever there
is a default in a given time step $ dt $, the probability for the
hidden variable $ Y $ to stay in state ``0" is zero, meaning it
jumps to state ``1'' (``risky state'') with (conditional)
probability 1\footnote{Note that the element $ F_{01}^{(i)} =
\lambda (N-i) $ specifies the probability of the joint transition $
P[(i,0) \rightarrow (i+1,1)]  = \lambda(N-i) dt$. On the other hand,
the marginal transition probability $ P[i \rightarrow i + 1] $ is
also equal to $ \lambda (N-i) $. Therefore, the conditional
probability $ P[ 0 \rightarrow 1|i \rightarrow i + 1] = 1 $ as
expected.} , or stays in state ``1'' with (conditional) probability
1 if it was already there at the beginning of the time step. The
matrix $ L^{(i)} $ is interpreted similarly.

In the BSLP model, the dynamics of $ Y $ is Markov as opposed to
being locally deterministic (as in the Davis-Lo model), thus
matrices $ L^{(i)} $ and $F^{(i)} $ will have very few, if any, zero
elements, as in general all transition probabilities between
different $Y$-states will be non-vanishing. Note that the generator
(\ref{2D}) can be identically re-written as follows: \bea
\label{2Da} A \, = \, \left( \begin{array}{clcrc}
\tilde{L}^{(0)}   & 0 & 0 & \cdots & 0  \\
0 & \tilde{L}^{(1)} &  0 & \cdots & 0 \\
0 &  0   & \tilde{L}^{(2)}  & \cdots & 0 \\
\vdots                        \\
 0  & 0  &  0 & \cdots & 0
\end{array} \right)
+
\left( \begin{array}{clcrc}
- \tilde{F}^{(0)}  &  F^{(0)} & 0 & \cdots & 0  \\
0 & - \tilde{F}^{(1)} &  F^{(1)} & \cdots & 0 \\
0 &  0   & - \tilde{F}^{(2)}  &  \cdots & 0 \\
\vdots                        \\
 0  & 0  &  0 & \cdots & 0
\end{array} \right)
\equiv A_0 + A_1 \nonumber
\eea where $ \tilde{L}^{(i)} =  L^{(i)} + \mbox{diag} \left( F^{(i)}
\mathbb{I} \right) $, $ \tilde{F}^{(i)} = \mbox{diag} \left( F^{(i)}
\mathbb{I} \right) $. Both matrices $ A_0 $ and $ A_1 $ can be
separately interpreted as generators, as in both of them
off-diagonal elements are positive, diagonal ones are negative, and
the row-wise sums are zeros. Note that they have  a straightforward
interpretation: generator $ A_0 $ corresponds to the case when no
$L$-transitions occur during the infinitesimal time step $ dt $,
while  $ A_1 $ allows
 jumps in the $L$-process during the
interval $ [t,t + dt] $, that are accompanied by possible
transitions between $Y$-states.
Note that this decomposition is a discrete counterpart
of~(\ref{formOfGenerator1}).

\appendix
\def\theequation{\thesection.\arabic{equation}}

\def\thesection{D}
\setcounter{equation}{0}

\section*{Appendix D: Adiabatic approximation in the continuous time BSLP model}

To get a tractable approximation to pricing index (or tranche)
options, we assume that the characteristic time scales of changes in
$Y$ and $ N$ ($ \tau_{spread} $  and $ \tau_{loss} $, respectively)
are significantly different. We expect this approximation to yield
accurate results because spreads change daily while loss states
change very infrequently. To acknowledge the presence of a hidden
small parameter $  \sim \tau_{spread}/ \tau_{loss} $ in the problem,
we re-write the infinitesimal generator (\ref{formOfGenerator1}) of
the BSLP model as follows: \beq \label{formOfGenerator} A =
\frac{1}{\varepsilon} \mathcal{L}_0 + \mathcal{L}_1 \eeq where the
parameter $ \varepsilon > 0 $ is assumed to be small.
Matrices $ \mathcal{L}_0 / \varepsilon $ and $ \mathcal{L}_1 $
correspond to parts of the generator responsible for the
$y$-dependent part (where the default counting variable $ N_t $ may
enter as a parameter) $ A_0 $ and a second term $A_1 $ that contains
both $ N_t $ and $ Y_t $ as dynamic variables. In particular, for
the OU specification (\ref{OUy}) with  mean reversion parameter $ a
= 1/ \varepsilon $ and variance $ \nu^2 = \sigma^2 /( 2 a ) $, we
obtain \beq \label{Lo} \mathcal{L}_0 = (\theta - y_0)
\frac{\partial}{\partial y_0} + \nu^2 \frac{
\partial^2 }{\partial y_0^2}
\eeq

Given these assumptions, we consider the adiabatic approximation
(see e.g.~\cite{Fouque} for financial applications of this method)
to find the forward and backward dynamics in the form of asymptotic
series in powers of $ \varepsilon $. We  restrict ourselves to a
calculation of the leading order adiabatic approximation for the
backward equation.

Consider the backward equation for some function $ B(x,t) $ of
initial values $ (x,t) $ (this function can be a conditional
expectation or a transition probability viewed as a function of the
initial variables):
\beq \label{backwardL} \frac{\partial
B}{\partial t} = - \mathcal{L} B = - \left( \frac{1}{ \varepsilon}
\mathcal{L}_0 + \mathcal{L}_1 \right) B \eeq
We look for a solution
in the form
\beq \label{epsExpansionBack} B = B_0 + \varepsilon B_1
+ \varepsilon^2 B_2 + \ldots \eeq
Plugging this into (\ref{backwardL}) and equating like powers of $
\varepsilon $, we obtain the chain of equations
 \bea \label{chain}
\mathcal{L}_0 B_0 &=& 0 \nonumber \\
\mathcal{L}_0 B_1 &=& - \frac{\partial B_0}{\partial t} - \mathcal{L}_1 B_0
\\
\mathcal{L}_0 B_2 &=& - \frac{\partial B_1}{\partial t} - \mathcal{L}_1 B_1
\nonumber
\eea
As the generator $ \mathcal{L}_0 $ depends only on $ y $, the
solution to the first equation is \beq \label{firstBack} B_0 = B_0
(x,t) \eeq i.e. a function (unknown at this stage) of $ x $ and $t$
only. To find this function, we multiply both sides of the second
equation in (\ref{chain}) by the stationary state $ \rho_0(y,x) $ of
the adjoint generator $ \mathcal{L}_0^{*} $, i.e. $
\mathcal{L}_0^{*} \rho_0(y,x) = 0 $. The left hand side of this
equation vanishes, hence the right hand side should vanish as well.
This implies that $ B_0(x,t) $ should solve the equation \beq
\label{B0} 0 = \la \frac{\partial B_0}{\partial t} + \mathcal{L}_1
B_0 \ra = \frac{\partial B_0}{\partial t} + \la \mathcal{L}_1 \ra
B_0 \eeq where \beq \label{L1average} \la \mathcal{L}_1 \ra \equiv
\int dy \rho_0(y,x) \mathcal{L}_1(y,x) \eeq That is, $ B_0(x,t) $
satisfies the backward equation in the $L$-space where the fast $
Y$-dynamics is integrated out in the generator $\mathcal{L}_1 $ with
weight equal to the invariant distribution $ \rho_0(y,x) $ of $
\mathcal{L}_0^{*} $. Correction terms $ B_1, \, B_2 , \ldots $ can
be found from further equations in (\ref{chain}).

Let us consider the leading order adiabatic approximation formulas.
Using (\ref{B0}) and (\ref{L1jump}), we obtain the following
equation for the lowest order term $ B_0(x,t) $: \beq \label{B02}
\frac{\partial B_0}{\partial t} = - \bar{\lambda}(x) \left( B_0(x +
\Delta x,t) - B_0(x,t) \right) \eeq where \beq \label{eff1D}
\bar{\lambda}(x) \equiv (1-x) q(x) f(x)  \la y \ra = (1-x) q(x) f(x)
\int dy \, y \rho_0(y,x) \eeq Thus to the leading order in the
adiabatic approximation, we find that the backward dynamics can be
described as the 1D pure default dynamics with a rescaled intensity.
If we further impose the condition that the intensity of this
effective 1D model should be equal to the intensity $ (1-x) \lambda
f(x) $ of the original default-only model, we obtain the following
expression for the drift adjustment factor $ q(x) $: \beq \label{qx}
q(x) = \frac{\lambda}{ \la y \ra_{x}} = \frac{\lambda}{\int y
\rho_0(y,x) dy} \eeq
One notes a similarity between(\ref{qx}) and
(\ref{driftConstraint}). The difference between them is that while
(\ref{driftConstraint}) contains the full time-dependent
distrubution $ \pi(Y_t,t| N_t) $, it is the steady state
distribution $ \rho_0(y,x) = \lim_{t \rightarrow \infty}
 \pi(Y_t,t| N_t)$ that appears in (\ref{qx}).

Corrections to the zero-order result $ B_0 = B_0 (x,t) $ can now be
found from equations (\ref{chain}). In particular, after the right
hand side of the second equation in (\ref{chain}) is ``centered''
(i.e. its expectation is zeroed by the choice of $ B_0 $), we can
write this equation as follows: \beq \label{Poisson} \mathcal{L}_0
B_1 = \left( \la \mathcal{L}_1 \ra - \mathcal{L}_1 \right) B_0 \eeq
This is the so-called Poisson equation that can be explicitly solved
to find the first correction $B_1 $, as described in \cite{Fouque}.

\jpmdisclaimer
\end{document}